\documentclass[]{article}

\usepackage[dvips]{epsfig}
\usepackage[]{amssymb}

\def\thebibliography#1{\section*{\normalsize \bf References 
 }\list
 {[\arabic{enumi}]}{\settowidth\labelwidth{[#1]}\leftmargin\labelwidth
 \advance\leftmargin\labelsep
 \usecounter{enumi}}
 \def\newblock{\hskip .11em plus .33em minus .07em}
 \sloppy\clubpenalty4000\widowpenalty4000
 \sfcode`\.=1000\relax}

\begin{document}

\sloppy

\twocolumn[

\begin{center} \large \bf 
  Optimization of alloy-analogy-based approaches \\
  to the infinite-dimensional Hubbard model 
\end{center}
\vspace{-3mm}

\begin{center} 
   M. Potthoff, T. Herrmann and W. Nolting
\end{center}
\vspace{-6mm}

\begin{center} \small \it 
   Lehrstuhl Festk\"orpertheorie,
   Institut f\"ur Physik, 
   Humboldt-Universit\"at zu Berlin, 
   D-10115 Berlin, 
   Germany
\end{center}
\vspace{2mm}

\begin{center}
\parbox{141mm}{ 
An analytical expression for the self-energy of the
infinite-dimensional Hubbard model is proposed that 
interpolates between different exactly solvable limits. 
We profit by the combination of two recent approaches 
that are based on the alloy-analogy (Hubbard-III) solution:
The modified alloy-analogy (MAA) which focuses on 
the strong-coupling regime, and the Edwards-Hertz approach 
(EHA) which correctly recovers the weak-coupling regime. 
Investigating the high-energy expansion of the EHA self-energy, it 
turns out that the EHA reproduces the first three exactly known 
moments of the spectral density only. This may be insufficient 
for the investigation of spontaneous magnetism. The analysis of 
the high-energy behavior of the CPA self-consistency equation 
allows for a new interpretation of the MAA: The MAA is the only 
(two-component) alloy-analogy that correctly takes into account the 
first four moments of the spectral density. For small $U$, however, 
the MAA does not reproduce Fermi-liquid properties. The defects 
of the MAA as well as of the EHA are avoided in the new approach.
We discuss the prospects of the theory and present numerical
results in comparison with essentially exact quantum Monte Carlo
data. The correct high-energy behavior of the self-energy is 
proved to be a decisive ingredient for a reliable description 
of spontaneous magnetism.

\vspace{4mm} 
}
\end{center}
\vspace{8mm} 
]

{\center \bf \noindent I. INTRODUCTION \\ \mbox{} \\} 

A central problem in solid-state physics concerns interacting 
electrons on a lattice. In dealing with itinerant magnetism,
heavy-fermion compounds or high-temperature superconductivity, 
for example, electron-correlation effects are of exceptional 
significance. Much insight into the fundamental role of electron 
correlations can be gained by studying the Hubbard model 
\cite{Hub63,Gut63,Kan63}. In spite of its apparent simplicity, 
an exact solution for the whole parameter range is not available 
up to now, and a completely satisfactory understanding of its 
properties has not yet been achieved.

A notable exception is the extreme case of the one-dimensional 
model. In particular, the exact solution for the ground state is 
known in $d=1$ \cite{LW68}. The opposite limit of high spatial 
dimensions $d$, which has been introduced by Metzner and Vollhardt
\cite{MV89}, is likewise important. For $d=\infty$ there are 
considerable simplifications that are due to the momentum 
independence of the electronic self-energy \cite{MH89b}. However, 
the infinite-dimensional model still remains non-trivial. It
is of special interest since its essential properties are 
expected to be comparable to those at low dimensions $d=2,3$. 
If an approximation scheme was available that 
is reliable for the entire range of the model parameters in
$d=\infty$, this would provide a proper dynamical mean-field 
theory in any dimension $d>1$ \cite{Vol93,GKKR96}.

Quantum Monte Carlo (QMC) \cite{Jar92,RZK92,GK92b} and exact 
diagonalization methods (ED) \cite{CK94,SRKR94} can yield 
essentially exact results but also suffer from limitations:
ED is restricted to a rather small number of 
orbitals, and thus a smooth density of states cannot be obtained. 
On the other hand, QMC yields its results for the 
discrete Matsubara energies or along the imaginary time axis.
Therefore, it is difficult to access the low-temperature regime
where statistical errors become important within the QMC method. 
Furthermore, to obtain dynamical quantities it becomes necessary
to perform an analytical continuation to the real axis which
constitutes an ill-conditioned numerical problem.
For these reasons the development of analytical (but approximate) 
methods for the 
infinite-dimensional Hubbard model still remains necessary. 

An analytical approach to the $d=\infty$ 
Hubbard model should be guided by rigorously solvable limiting 
cases. There are a number of exact results which impose strong
necessary conditions on any approximation: 
(i) Second-order perturbation 
theory around the Hartree-Fock solution \cite{SC91} (SOPT-HF) or 
self-consistent SOPT \cite{MH89b,HC94} yield the correct self-energy 
in the weak-coupling regime.
(ii) The exact low-energy (Fermi-liquid) properties of the 
self-energy 
and the Luttinger sum rule \cite{Lut60} in particular
have been analyzed in Ref.\ \cite{MH89b} for the $d=\infty$ case.
(iii) The atomic limit of vanishing hopping $t=0$ represents an 
extreme but important limit. The first non-trivial correction to
the atomic limit can be obtained 
from a canonical transformation of the Hubbard model 
that has first been considered by Harris and Lange 
\cite{HL67,EOMS94}. 
For $U\mapsto \infty$ the
approach provides rigorous information on the centers of 
gravity as well as on the spectral weights of the lower and upper 
Hubbard bands. 
(iv) The high-energy behavior of the self-energy is determined
by the moments of the spectral density which can be calculated
independently. For any $U$ the high-energy behavior is 
important for a qualitatively correct description of the 
charge-excitation peaks. 
For $U\mapsto \infty$ the first four moments are 
necessary to be consistent with the results of Harris and Lange
\cite{PWN97}.
(v) Finally, the Falicov-Kimball model (FKM) \cite{FK69} can be
considered as a special limit of the Hubbard model. It is
obtained by formally 
introducing a spin-dependent hopping and allowing only one of the 
two spin species to be mobile. The $d=\infty$ FKM is exactly solvable
as has first been shown by Brandt and Mielsch \cite{BM}.
The self-energy is given by the Hubbard-III alloy-analogy (AA) 
solution \cite{Hub64b}.

It may be interesting to have a theory at one's disposal
that for arbitrary fillings recovers the 
weak-coupling, the atomic and the FKM limit, 
that is able to reproduce the first four moments
of the spectral density and thereby the exact strong-coupling
results of Harris and Lange, and that allows to directly treat 
zero temperature and real 
energies. The construction of such an approach is the main 
purpose of the present paper.
To make contact with the FKM, the approach will be based on the
alloy-analogy solution. The AA 
represents an attractive approximation which is simple, 
self-consistent and non-perturbative.
On the other hand, there are defects that show up in other limits
and that have given rise to modifications of the conventional AA:

A mean-field 
theory has been constructed in Refs.\ \cite{JV93,JMV93} which is 
conceptually similar to the AA but in contrast to the latter 
derivable from an explicit free-energy functional and thus 
thermodynamically consistent and of variational character.
This approach as well as the AA itself, however, cannot reproduce
the aforementioned weak- and strong-coupling limits. 
The weak-coupling behavior
of the AA is corrected within the Edwards-Hertz approach
(EHA) \cite{EH90,Edw93,WC94,WC95}. 
A drawback of the EHA and also of the AA, however, consists in the
fact that the strong-coupling results of Harris and Lange cannot
be recovered.
A recently proposed modified alloy analogy (MAA) \cite{HN96} improves
upon the AA and ensures consistency with the results of Harris and 
Lange for $U\mapsto \infty$. On the other hand, the MAA is not able
to reproduce the Fermi-liquid properties for small $U$.

Our study aims at a proper combination of the MAA and the EHA which
keeps their advantages and avoids their defects.
After the necessary theoretical preparations (Sec.~II), 
we derive a helpful alternative interpretation of the MAA in 
Sec.~III. Sec.~IV concerns the high-energy expansion of the EHA 
self-energy. The analysis of the MAA and the EHA is a necessary
condition for the concise presentation of the combined theory in 
Sec.~V. Numerical results obtained within the new method will be 
discussed for the paramagetic phase and in particular with respect 
to ferromagnetism (Sec.~VI) before we come 
to the conclusions in Sec.~VII.
\\

{\center \bf \noindent II. PREPARATIONS \\ \mbox{} \\} 

Using standard notations, the Hubbard Hamiltonian reads:
\begin{equation}
  H = \sum_{ij\sigma} \left( T_{ij} - \mu \delta_{ij} \right)
  c^\dagger_{i\sigma} c_{j\sigma} 
  + \frac{1}{2} U \sum_{i \sigma} n_{i\sigma} n_{i -\sigma} \: .
\label{eq:hubbard}
\end{equation}
The limit $d\mapsto \infty$ has 
to be taken with the proper scaling of the hopping integral between 
nearest neighbors to ensure the model to remain non-trivial 
\cite{MV89}. For a hyper-cubic lattice with coordination number
$Z=2d$ and hopping
$T_{\langle ij \rangle}\equiv t=t^\ast/\sqrt{2Z}$
($t^\ast=\mbox{const.}$) we get a Gaussian for the (non-interacting)
Bloch-density of states (BDOS) \cite{MV89}:
\begin{equation}
  \rho^{\rm (B)}(E) = \frac{1}{t^\ast \sqrt{\pi}} 
  e^{ -\left( E/t^\ast \right)^2 } \: .
\label{eq:bdoshc}
\end{equation}
A generalization of the $d=3$ fcc lattice to infinite dimensions 
with $Z=2d(d-1)$ results in a strongly 
asymmetric Bloch-density of states \cite{MH91}:
\begin{equation}
  \rho^{\rm (B)}(E) = \frac{e^{-(1+\sqrt{2}E/t^\ast)/2}}
  {t^\ast \sqrt{\pi (1+\sqrt{2}E/t^\ast)}}
  \: ,
\label{eq:bdosfcc}
\end{equation}
where the hopping integral between nearest neighbors $T_{ij}$ is 
positive and $t=t^\ast/\sqrt{Z}$ \cite{tastdisc}.

In infinite dimensions the self-energy $\Sigma_{\sigma}(E)$ is 
$\bf k$ independent or site-diagonal \cite{MH89b}. For a 
homogeneous phase, the on-site one-electron Green function thus 
depends on the lattice geometry via the BDOS only:
\begin{equation}
  G_\sigma(E) = \int_{-\infty}^\infty \frac{\hbar 
  \rho^{\rm (B)}(z)}
  {E-(z-\mu) - \Sigma_{\sigma}(E)} \: dz \: .
\label{eq:grhm}
\end{equation}
Let us consider its high-energy expansion:
\begin{equation}
  G_{\sigma}(E) = \hbar \sum_{m=0}^\infty 
  \frac{M_{\sigma}^{(m)}}{E^{m+1}} \: .
\label{eq:gexp}
\end{equation}
The expansion coefficient are the moments
\begin{equation}
  M^{(m)}_{\sigma} = \frac{1}{\hbar} \int_{-\infty}^\infty E^m
  A_{\sigma}(E) \, dE 
\label{eq:momentsdef}
\end{equation}
of the spectral density
$A_{\sigma}(E) = -\frac{1}{\pi} \mbox{Im} \, G_{\sigma}(E+i0^+)$. 
They can be calculated exactly by making use of an 
alternative but equivalent representation:
\begin{equation}
  M^{(m)}_{\sigma} = \langle [ {\cal L}^m c_{i\sigma} , 
  c_{i\sigma}^\dagger ]_+ \rangle \: .
\label{eq:moments}
\end{equation}
Here ${\cal L O}=[{\cal O},H]_-$ denotes the commutator of an 
operator $\cal O$ with the Hamiltonian, and $[\cdots , \cdots]_+$ 
is the anticommutator. The calculation along Eq.\ (\ref{eq:moments})
is straightforward. However, with increasing order $m$ the moments
include equal-time correlation functions of higher and higher 
order which have to be expressed by some means in terms of known 
quantities again. This fact limits the number of moments that 
can be used in practice. For the Hubbard model the moments are 
useful up to $m=3$ (see Refs.\ \cite{NB89,PN96}, for example). In 
particular, we obtain for $d=\infty$:
\begin{eqnarray}
  M^{(0)}_{\sigma} \!\!\! & = & \!\!\! 1 \: ,
  \nonumber \\
  M^{(1)}_{\sigma} \!\!\! & = & \!\!\! 
  \widetilde{T}_0 + U n_{-\sigma} \: ,
  \nonumber \\
  M^{(2)}_{\sigma} \!\!\! & = &\!\!\! 
  \sum_j \widetilde{T}_{ij} \widetilde{T}_{ji} 
  + 2 \widetilde{T}_0 U 
  n_{-\sigma} + U^2 n_{-\sigma} \: ,
  \nonumber \\
  M^{(3)}_{\sigma} \!\!\! & = & \!\!\! 
  \sum_{jk} \widetilde{T}_{ij} \widetilde{T}_{jk} \widetilde{T}_{ki}
   + 3 U n_{-\sigma} \sum_j \widetilde{T}_{ij} \widetilde{T}_{ji}
  \nonumber \\ \!\!\! & + & \!\!\! 
  \widetilde{T}_0 U^2 n_{-\sigma} ( 2 + n_{-\sigma} )
  + U^3 n_{-\sigma} 
  \nonumber \\ \!\!\! & + & \!\!\! 
  U^2 n_{-\sigma} (1- n_{-\sigma} )
  \widetilde{B}_{-\sigma} \: .
\label{eq:mom}
\end{eqnarray}
For abbreviation we have defined:
$\widetilde{T}_{ij} = T_{ij} - \mu \delta_{ij}$, $\widetilde{T}_0
=\widetilde{T}_{ii}$ and
$\widetilde{B}_{\sigma} = B_{\sigma} - \mu$. This so-called
`band-shift' consists of higher-order correlation functions,
\begin{equation}
  {B}_{\sigma} = {T}_0 +
  \frac{1}{ n_{\sigma} ( 1 - n_{\sigma} ) }
  \sum_{j\ne i} {T}_{ij} 
  \langle c^\dagger_{i\sigma} c_{j\sigma} 
  (2 n_{i-\sigma} - 1) \rangle 
  \: ,
\label{eq:bfin}
\end{equation}
but can be expressed in terms of the 
one-particle Green function \cite{PN96}.
Note that while the hopping $T_{ij}$ as well as the correlation
functions $\langle c^\dagger_{i\sigma}c_{j\sigma}(2n_{i-\sigma}-1) 
\rangle$ scale as $1/\sqrt{d}$ for $d\mapsto \infty$, the lattice sum 
in (\ref{eq:bfin}) remains finite. Contrary to the finite-dimensional
case, however, the band-shift is $\bf k$ independent for $d=\infty$.

The coefficients in the $1/E$ expansion of the self-energy,
\begin{equation}
  \Sigma_{\sigma}(E) = \sum_{m=0}^\infty 
  \frac{C^{(m)}_{\sigma}}{E^{m}} \: ,
\label{eq:sexp}
\end{equation}
are obtained by inserting Eqs.\ (\ref{eq:gexp}) and 
(\ref{eq:sexp}) into (\ref{eq:grhm}) and taking the moments of 
the BDOS from (\ref{eq:mom}) for $U=0$:
\begin{eqnarray}
  C^{(0)}_{\sigma} \!\!\! & = & \!\!\! U n_{-\sigma} \: ,
  \nonumber \\
  C^{(1)}_{\sigma} \!\!\! & = & \!\!\! U^2 n_{-\sigma} 
  \left( 1 - n_{-\sigma} \right) \: ,
  \nonumber \\
  C^{(2)}_{\sigma} \!\!\! & = & \!\!\! U^2 n_{-\sigma}
  \left( 1 - n_{-\sigma} \right)
  \left( \widetilde{B}_{-\sigma} 
  + U (1- n_{-\sigma}) \right) \: .
  \nonumber \\
\label{eq:smom}
\end{eqnarray}

Let us mention some available rigorous results for the weak- and 
the strong-coupling regime. For $U\mapsto \infty$ the spectrum is
dominated by the two charge-excitation peaks (Hubbard bands).
At each $\bf k$ point the centers of gravity as well as the 
spectral weights of the lower and the upper Hubbard band can
be calculated exactly within a perturbational approach due to
Harris and Lange \cite{HL67,EOMS94} that uses $t/U$
as an expansion parameter. For $d=\infty$ we have:
\begin{eqnarray}
  T^{\rm (HL)}_{1\sigma}({\bf k}) \!\!\! & = & \!\!\! 
  (1-n_{-\sigma}) \epsilon({\bf k}) + n_{-\sigma} B_{-\sigma}
  + {\cal O}(t/U) \: ,
  \nonumber \\
  T^{\rm (HL)}_{2\sigma}({\bf k}) \!\!\! & = & \!\!\! 
  U + n_{-\sigma} \epsilon({\bf k}) + (1-n_{-\sigma}) B_{-\sigma}
  + {\cal O}(t/U) \: ,
  \nonumber \\
  \alpha^{\rm (HL)}_{1\sigma}({\bf k}) \!\!\! & = & \!\!\! 
  1 - \alpha^{\rm (HL)}_{2\sigma}({\bf k}) 
  = 1-n_{-\sigma} + {\cal O}(t/U)
  \: .
\label{eq:hl}
\end{eqnarray}
The only difference to the finite-dimensional case \cite{HL67} 
consists in the fact that the $\bf k$ dependence of the centers of
gravity $T^{\rm (HL)}_{p\sigma}({\bf k})$ and the weights
$\alpha^{\rm (HL)}_{p\sigma}({\bf k})$ ($p=1,2$) is exclusively due 
to the Bloch dispersion $\epsilon({\bf k})$. For finite $d$ an 
additional $\bf k$ dependence is introduced via the band shift.

In the weak-coupling regime
the usual perturbative approach applies. Up to order
$U^2$ the self-energy is given by:
\begin{equation}
  \Sigma_{\sigma}^{\rm (SOPT)}(E) = U n_{-\sigma} + 
  \Sigma^{\rm (SOC)}_{\sigma}(E) \: .
\label{eq:sigmaipt}
\end{equation}
The first-order term is the Hartree-Fock self-energy; the 
second-order contribution reads:
\begin{eqnarray}
  && \hspace{-9mm} \Sigma^{\rm (SOC)}_{\sigma}(E) = 
  \frac{U^2}{\hbar^3} 
  \! \int \!\!\! \int \!\!\! \int 
  \frac{A^{(1)}_{\sigma}(x) A^{(1)}_{-\sigma}(y) 
  A^{(1)}_{-\sigma}(z)}{E-x+y-z} \times
  \nonumber \\ &&
  (f(x) f(-y) f(z) + f(-x) f(y) f(-z))
  \: dx \, dy \, dz \: .
\label{eq:sigma2}
\end{eqnarray}
Here $f(x)$ is the Fermi function, and $A^{(1)}_\sigma(E)$ is defined 
as the free ($U=0$) spectral density shifted in energy by a (possibly 
spin-dependent) constant $E_\sigma$:
\begin{equation}
  A^{(1)}_{\sigma}(E) = A^{(0)}_{\sigma}(E-E_\sigma) \: ,
\label{eq:rho1}
\end{equation}
The shifts $E_\sigma$ are introduced for
later use. With $E_\sigma=0$ the plain or conventional second-order
perturbation theory (SOPT) and with $E_\sigma=U n_{-\sigma}$ the 
SOPT around the Hartree-Fock solution \cite{SC91} is recovered.
Both versions of SOPT as well as the self-consistent SOPT 
\cite{MH89b} are identical up to order $U^2$.
\\

{\center \bf \noindent III. MODIFIED ALLOY-ANALOGY \\ \mbox{} \\} 

The original alloy-analogy solution (AA) by Hubbard \cite{Hub64b} 
is at
variance with both, the exact weak- and strong-coupling results.
Recently, a modified alloy-analogy solution (MAA) has been proposed
\cite{HN96} to overcome the drawbacks of the conventional 
AA with
respect to the limit $U\mapsto \infty$.
Any two-component alloy analogy requires the specification 
of the two atomic levels and the corresponding concentrations. 
The self-energy is then obtained from the CPA equation \cite{VKE68}:
\begin{equation}
  0 = \sum_{p=1}^2 x_{p\sigma} \:
  \frac{E_{p\sigma} - \Sigma_\sigma(E) - T_0}
  {1 - \frac{1}{\hbar} G_\sigma(E) \:
  [E_{p\sigma} - \Sigma_\sigma(E) -T_0] } \: .
\label{eq:cpaeq}
\end{equation}
Within the conventional AA the levels and weights 
are taken from the atomic-limit solution \cite{Hub63}, i.~e.:
$E_{1\sigma}^{\rm (AA)} = T_0 \: , 
E_{2\sigma}^{\rm (AA)} = T_0 + U$ and
$x_{1\sigma}^{\rm (AA)} = 1 - n_{-\sigma} \: ,
x_{2\sigma}^{\rm (AA)} = n_{-\sigma}$.
This choice, 
however, is by no means predetermined. Within the MAA \cite{HN96} the 
self-energy $\Sigma_\sigma^{\rm (MAA)}(E)$ is still 
obtained from the CPA equation (\ref{eq:cpaeq}), but the 
atomic levels and the concentrations are replaced by:
\begin{eqnarray}
  E_{p\sigma}^{\rm (MAA)} \!\!\!\! & = & \!\!\!\! \frac{1}{2} 
  \left( T_0 + U + B_{-\sigma} \right) + (-1)^p \times
  \nonumber \\ && \hspace{-8mm}
  \sqrt{
  \frac{1}{4} \left( U + B_{-\sigma} - T_0 \right)^2 
  + U n_{-\sigma} \left( T_0 - B_{-\sigma} \right) 
  } \: ,
  \nonumber \\ &&
\label{eq:maalevels}
\end{eqnarray}
and
\begin{equation}
  x_{1\sigma}^{\rm (MAA)} = 
  \frac{B_{-\sigma} + U (1 - n_{-\sigma}) - E_{1\sigma}^{\rm (MAA)}}
  {E_{2\sigma}^{\rm (MAA)} - E_{1\sigma}^{\rm (MAA)}}
  = 1 - x_{2\sigma}^{\rm (MAA)} \: .
\label{eq:maacons}
\end{equation}
The expressions include the band shift $B_{-\sigma}$. If 
$B_{-\sigma}$ is replaced by $T_0$, the MAA reduces
to the conventional AA. As can be seen from Eq.\ (\ref{eq:bfin}) 
this replacement
is correct for the atomic and for the ``FKM limit''. Furthermore,
$B_{-\sigma} = T_0$ is required by particle-hole symmetry 
in the case of a paramagnet at half-filling. 
In all other cases the MAA is different from the AA. In 
particular, via the band shift the atomic levels might now
become spin-dependent. 

In the original work \cite{HN96} the MAA has been derived by 
referring to the spectral-density approach (SDA) 
\cite{NB89,PN96,HN97b}. Considering the so-called split-band 
regime of the CPA \cite{VKE68}, the consistency of the MAA with 
the results of Harris and Lange (\ref{eq:hl}) could be proven 
for $U\mapsto \infty$. 

For later purposes it is important to give another interpretation 
of the MAA which is different from the original one in Ref.\ 
\cite{HN96}: Since the atomic values and the concentrations 
that are needed to construct an alloy analogy for the Hubbard 
model are not predetermined, one can ask for the optimal choice 
(in some sense). In the following we argue that the MAA is obtained 
if one demands that the resulting theory should correctly take into
account the first four moments of the spectral density given 
in Eq.\ (\ref{eq:mom}):

We consider the CPA equation (\ref{eq:cpaeq}) as an equation
for the self-energy that contains four unknown parameters,
$E_{p\sigma}$ and $x_{p\sigma}$ ($p=1,2$), which have to be fixed
by imposing the correct results for the moments of the spectral
density. Since the moments are closely related to the high-energy 
behavior of the Green function, we have to expand the CPA equation 
in powers of $1/E$ using Eqs.\ (\ref{eq:gexp}) and (\ref{eq:sexp}).
Considering terms up to $1/E^2$, this yields the following 
equations:
\begin{eqnarray}
  1 \!\!\! & = & \!\!\!
  \sum_p x_{p\sigma} \: ,
  \nonumber \\
  0 \!\!\! & = & \!\!\!
  \sum_p x_{p\sigma} \: 
  (E_{p\sigma} - T_0 - C_\sigma^{(0)}) \: ,
  \nonumber \\
  0 \!\!\! & = & \!\!\! \sum_p x_{p\sigma} \:
  \Big[ (E_{p\sigma} - T_0 - C_\sigma^{(0)})^2 M_\sigma^{(0)}
  - C_\sigma^{(1)}
  \Big] \: ,
  \nonumber \\
  0 \!\!\! & = & \!\!\!
  \sum_p x_{p\sigma} \: 
  \Big[ (E_{p\sigma} - T_0 - C_\sigma^{(0)})^3 (M_\sigma^{(0)})^2
  \nonumber \\
  && + \;
  (E_{p\sigma} - T_0 - C_\sigma^{(0)})^2 M_\sigma^{(1)}
  \nonumber \\
  && - \;
  2 (E_{p\sigma} - T_0 - C_\sigma^{(0)})
  C_\sigma^{(1)} M_\sigma^{(0)} - C_\sigma^{(2)}
  \Big] \: .
\label{eq:cpahigh}
\end{eqnarray}
Inserting the exact expansion coefficients (\ref{eq:mom}) and
(\ref{eq:smom}) for the Green function and for the self-energy
and solving for $E_{p\sigma}$ and $x_{p\sigma}$, indeed yields 
the MAA results (\ref{eq:maalevels}) and (\ref{eq:maacons}).

This proves that the MAA not only recovers the exact strong-coupling
results of Harris and Lange, but also (for arbitrary $U$) respects 
the conditions that are imposed on the overall shape of the spectral
density by their exactly known first four moments. In this sense
the MAA can be termed to be an optimal alloy analogy.

It is not very difficult to analyze the high-energy behavior of 
the conventional AA. Starting from Eq.\ (\ref{eq:cpahigh}) and 
inserting the AA levels and concentrations, 
it shows up that the last equation cannot
be satisfied. While the AA yields the correct coefficients up to
order $1/E$ in the expansion of the self-energy, it turns out that
the coefficient $C_\sigma^{(2)}$ is not recovered. Instead, one 
gets: $C_\sigma^{(2, {\rm AA)}} = C_\sigma^{(2)} |_{B_{-\sigma} 
\mapsto T_0}$. This is only exact in the atomic limit. Equivalently,
it can be stated that the AA recovers three (instead of four)
moments of the spectral density only.

While the MAA (in the sense of Harris and Lange) is correct in the 
strong-correlation regime, there is a severe drawback of the method:
It fails to reproduce the Fermi-liquid properties for small 
interactions $U$. The same defect is inherent in the conventional
AA. 
\\

{\center \bf \noindent IV. EDWARDS-HERTZ APPROXIMATION \\ \mbox{} \\}

A correction of the conventional AA with respect to Luttinger's
requirements and Fermi-liquid properties is due to Edwards 
and Hertz \cite{EH90}. Just as the MAA improves upon the 
strong-coupling regime of the AA, the Edwards-Hertz approximation 
(EHA) corrects the AA in the weak-coupling regime. 

The self-energy within the Edwards-Hertz approximation (EHA) is 
implicitly given by \cite{EH90} (see also Refs.\ \cite{Edw93,WC94}):
\begin{equation}
  \Sigma_\sigma^{\rm (EHA)}(E) = \frac{U n_{-\sigma}}{
  1 - \frac{1}{\hbar} \widetilde{G}_\sigma(E) \:
  ( U - \Sigma^{\rm (EHA)}_\sigma(E) )} \: .
\label{eq:sigeha}
\end{equation}
This equation is formally equivalent to the CPA equation 
(\ref{eq:cpaeq}) provided that the atomic levels and the 
concentrations are taken to be those of the conventional AA. 
Contrary to the AA,
however, the one-particle Green function that occurs in the 
CPA equation, $G_\sigma(E)$, has been replaced by:
\begin{equation}
  \widetilde{G}_\sigma(E) = 
  \frac{\hbar}{U^2 n_{-\sigma} (1-n_{-\sigma})} \:
  \Sigma^{\rm (SOC)}_\sigma
  (E-\Sigma^{\rm (EHA)}_\sigma(E)+E_\sigma) \: .
\label{eq:mprop}
\end{equation}
$\Sigma^{\rm (SOC)}_\sigma(E)$ is the second-order contribution to 
the SOPT self-energy introduced in Eq.\ (\ref{eq:sigma2}).
The replacement is more or less ad hoc to enforce the correct
weak-coupling behavior of the theory. 

The shifts $E_\sigma$ that appear in the definition (\ref{eq:rho1})
of the spectral density $A_\sigma^{(1)}(E)$ and once more in Eq.\ 
(\ref{eq:mprop}) have been introduced by Wermbter and Czycholl 
\cite{WC94}. They are necessary to ensure that the EHA is exact 
in the atomic limit for all band-fillings. The shifts have to be 
determined self-consistently by the following requirement:
\begin{equation}
  n_\sigma = \int_{-\infty}^\infty f(E) \, A_\sigma(E) \, dE 
  = \int_{-\infty}^\infty f(E) \, A^{(1)}_\sigma(E) \, dE \: .
\label{eq:nnhf}
\end{equation}

The EHA should be regarded as an interpolation scheme. It yields
the correct results in the atomic as well as in the FKM limit and 
is exact for small $U$ up to order $U^2$ \cite{WC94}. The latter 
implies that the expansion in $U$ satisfies Luttinger's 
requirements. However, Fermi-liquid behavior breaks down 
in a parameter region where the series diverges \cite{EH90}.

Let us now investigate the high-energy behavior of the EHA
self-energy:
\begin{equation}
  \Sigma^{\rm (EHA)}_{\sigma}(E) = \sum_{m=0}^\infty 
  \frac{C^{(m, {\rm EHA)}}_{\sigma}}{E^{m}}  \: .
\label{eq:sehaexp}
\end{equation}
First, we need the high-energy expansion of the modified propagator
$\widetilde{G}_\sigma(E)$. Applying the Kramers-Kronig relation, it 
can be written in the form:
\begin{eqnarray}
  \widetilde{G}_\sigma(E) \!\!\! & = & \!\!\!
  \frac{\hbar}{U^2 n_{-\sigma}(1-n_{-\sigma})} \times
  \nonumber \\ && \hspace{-5mm}
  \int \frac{-\frac{1}{\pi} \mbox{Im} \: 
  \Sigma^{\rm (SOC)}_\sigma(E'+i0^+)}
  {E-\Sigma^{\rm (EHA)}_\sigma(E)+E_\sigma - E'} \: dE' \: .
\end{eqnarray}
When expanding in powers of $1/E$, 
the following relations are needed:
\begin{eqnarray}
  \int - \frac{1}{\pi} \mbox{Im} \: 
  \Sigma^{\rm (SOC)}_\sigma(E+i0^+) \: dE
  \!\!\! & = & \!\!\!
  U^2 n_{-\sigma} (1-n_{-\sigma}) \: ,
  \nonumber \\
  \int - \frac{1}{\pi} \mbox{Im} \: 
  \Sigma^{\rm (SOC)}_\sigma(E+i0^+) \: E \: dE
  \!\!\! & = & \!\!\! \nonumber \\ && \hspace{-30mm}
  U^2 n_{-\sigma} (1-n_{-\sigma}) \: \left(
  B^{(1)}_{-\sigma} - \mu + E_{\sigma} \right) \: .
  \nonumber \\
\end{eqnarray}
Using Eq.\ (\ref{eq:nnhf}) these results can be derived from 
Eq.\ (\ref{eq:sigma2}) immediately. For abbreviation we have
defined:
\begin{equation}
  {B}^{(1)}_{\sigma} = {T}_0 +
  \frac{ 2 n_{-\sigma} - 1 }{ n_{\sigma} ( 1 - n_{\sigma} ) }
  \sum_{j\ne i} \widetilde{T}_{ij} 
  \langle c^\dagger_{i\sigma} c_{j\sigma} \rangle^{(1)}
  \: ,
\label{eq:b1def}
\end{equation}
where the one-particle correlation function results from the free 
($U=0$) but shifted off-diagonal Green function:
$\langle c^\dagger_{i\sigma} c_{j\sigma} \rangle^{(1)} =
-\frac{1}{\hbar \pi} \int f(E) \: \mbox{Im} \:
G_{ij\sigma}^{(0)}(E+i0^+-E_\sigma) dE$. 
The expansion of the modified propagator then reads:
\begin{equation}
  \frac{1}{\hbar} \widetilde{G}_\sigma(E) = 
  \frac{1}{E} +  
  \left( B^{(1)}_{-\sigma} - \mu + U n_{-\sigma} \right) 
  \frac{1}{E^2} + \cdots
  \: .
\label{eq:gtilexp}
\end{equation}
The coefficients in this expansion are the moments of the (fictive)
spectral density $\widetilde{A}_\sigma(E) \equiv - \frac{1}{\pi}
\mbox{Im} \, \widetilde{G}_\sigma(E+i0^+)$ (This will become 
important in the next section). Inserting the result and the
ansatz (\ref{eq:sehaexp}) in Eq.\ (\ref{eq:sigeha}) and
grouping the terms in powers of $1/E$, we finally obtain:
\begin{eqnarray}
  C^{(0, {\rm EHA)}}_{\sigma} \!\!\! & = & \!\!\! U n_{-\sigma} \: ,
  \nonumber \\
  C^{(1, {\rm EHA)}}_{\sigma} \!\!\! & = & \!\!\! U^2 n_{-\sigma} 
  \left( 1 - n_{-\sigma} \right) \: ,
  \nonumber \\
  C^{(2, {\rm EHA)}}_{\sigma} \!\!\! & = & \!\!\! U^2 n_{-\sigma}
  \left( 1 - n_{-\sigma} \right) \times
  \nonumber \\ && \hspace{-4mm}
  \left( {B}^{(1)}_{-\sigma} - \mu 
  + U (1- n_{-\sigma}) \right) \: .
\label{eq:smomeha}
\end{eqnarray}

Comparing with Eq.\ (\ref{eq:smom}), we notice that the first
two coefficients $C^{(0)}_{\sigma}$ and $C^{(1)}_{\sigma}$ are
predicted correctly. As in the conventional AA, however, the third
expansion coefficient turns out to be wrong: 
$C_\sigma^{(2, {\rm EHA)}} = C_\sigma^{(2)} |_{B_{-\sigma} 
\mapsto {B}^{(1)}_{-\sigma}}$. Again, the replacement 
$B_{-\sigma} \mapsto {B}^{(1)}_{-\sigma}$ is correct
in the atomic limit only. Furthermore, via Eqs.\ (\ref{eq:grhm}), 
(\ref{eq:gexp}) and (\ref{eq:sexp}) we can 
conclude that the fourth moment of the spectral density, 
$M_\sigma^{(3)}$, is not recovered within the EHA. Finally,
this implies that for $U\mapsto \infty$ the EHA is at variance
with the exact results of Harris and Lange [Eq.\ (\ref{eq:hl})].
\\

{\center \bf \noindent V. COMBINING THE APPROACHES \\ \mbox{} \\} 

We are now in a position to profit from the preceding analysis 
of the MAA and the EHA. We like to tackle the question whether a
proper combination of both approaches is able to avoid their
respective defects and to keep their advantages.

The following simple idea suggests itself: We start from the 
general CPA equation (\ref{eq:cpaeq}), replace the one-particle
Green function $G_\sigma(E)$ by the modified propagator 
$\widetilde{G}_\sigma(E)$
as in the EHA, but take the atomic levels $E_{p\sigma}$ and 
the weights $x_{p\sigma}$ from the MAA according to Eqs.\ 
(\ref{eq:maalevels}) and (\ref{eq:maacons}). It can be shown,
however, that such an ansatz is neither correct up to order
$U^2$ nor it recovers the high-energy coefficients up to
order $1/E^2$.

We thus have to disregard the MAA levels and weights, but can try 
to retain the basic concept of the approach. In Sec.\ IV we have 
interpreted the MAA as being the only (two-component) alloy 
analogy that correctly takes into account the first four moments of 
the spectral density: Insisting on the correctness of the moments,
unambiguously determines the levels and weights that appear in 
the CPA equation. Therefore, if we change the CPA equation by 
introducing the modified propagator $\widetilde{G}_\sigma(E)$, 
we must allow the parameters to adjust in order
to keep the moments correct.
This concept turns out to be successful indeed. For clarity in the 
notations let us refer to this approach as an ``interpolating 
alloy-analogy-based'' approximation (IAA) since we can show that 
it recovers the weak-coupling limit and satisfies the requirements 
of Harris and Lange for strong $U$.

To begin with, we once more consider Eq.\ (\ref{eq:cpaeq}) with
${G}_\sigma(E)$ replaced by $\widetilde{G}_\sigma(E)$ and with
a priori unknown parameters $E_{p\sigma}$ and $x_{p\sigma}$. As 
in the EHA the modified propagator $\widetilde{G}_\sigma(E)$ is 
defined via Eq.\ (\ref{eq:mprop}). To make contact with the
moments of the spectral density, we have to expand the equation
in powers of $1/E$. This yields Eq.\ (\ref{eq:cpahigh}) where 
the quantities $M^{(m)}_\sigma$ now have to be interpreted as the 
moments of the modified propagator $\widetilde{G}_\sigma(E)$.
Fortunately, only the first two moments are needed which can be
taken from Eq.\ (\ref{eq:gtilexp}). To ensure the correctness of
the moments (\ref{eq:mom}) 
within the IAA, we then have to insert the exact 
expressions (\ref{eq:smom}) for the coefficients $C_\sigma^{(m)}$,
which yields the following system of equations:
\begin{eqnarray}
&&  \sum_p x_{p\sigma}^{\rm (IAA)}
  = 1 \,
  \nonumber \\
&&  \sum_p x_{p\sigma}^{\rm (IAA)} \: 
  (E_{p\sigma}^{\rm (IAA)} - T_0)
  =  U n_{-\sigma} \: ,
  \nonumber \\
&&  \sum_p x_{p\sigma}^{\rm (IAA)} \: 
  (E_{p\sigma}^{\rm (IAA)} - T_0)^2
  =  U^2 n_{-\sigma} \: ,
  \nonumber \\
&&  \sum_p x_{p\sigma}^{\rm (IAA)} \: 
  (E_{p\sigma}^{\rm (IAA)} - T_0)^3
  =  U^3 n_{-\sigma}
  \nonumber \\ 
&& \hspace{5mm}
  + \: U^2 n_{-\sigma} (1-n_{-\sigma}) 
  \left( B_{-\sigma} - {B}_{-\sigma}^{(1)} \right) \: .
\label{eq:mehahigh}
\end{eqnarray}
Solving for $E_{p\sigma}^{\rm (IAA)}$ and $x_{p\sigma}^{\rm (IAA)}$,
we obtain:
\begin{eqnarray}
  E_{p\sigma}^{\rm (IAA)} \!\!\!\! & = & \!\!\!\! T_0 + \frac{1}{2} 
  \left( U + B_{-\sigma} - B^{(1)}_{-\sigma} \right) + (-1)^p \times
  \nonumber \\ && \hspace{-8mm}
  \sqrt{
  \frac{1}{4} \left( U + B_{-\sigma} - B^{(1)}_{-\sigma} \right)^2 
  + U n_{-\sigma} \left( B^{(1)}_{-\sigma} - B_{-\sigma} \right) 
  } \: ,
  \nonumber \\ &&
\label{eq:iaalevels}
\end{eqnarray}
and
\begin{eqnarray}
  x_{1\sigma}^{\rm (IAA)} \!\!\! & = & \!\!\!
  \frac{B_{-\sigma} - B^{(1)}_{-\sigma} + T_0
  + U (1 - n_{-\sigma}) - E_{1\sigma}^{\rm (IAA)}}
  {E_{2\sigma}^{\rm (IAA)} - E_{1\sigma}^{\rm (IAA)}} 
  \nonumber \\ 
  \!\!\! & = & \!\!\! 1 - x_{2\sigma}^{\rm (IAA)} \: .
\label{eq:iaacons}
\end{eqnarray}
The results turn out to represent a slight modification of the 
corresponding MAA results which are obtained if $B^{(1)}_{-\sigma}$ 
is (ad hoc) replaced by $T_0$. In the last section we have 
recognized that the more simple one-particle correlation functions 
$B^{(1)}_{-\sigma}$ instead of the higher-order correlations 
$B_{-\sigma}$ appear in the expansion coefficient 
$C_\sigma^{(2, {\rm EHA)}}$ of the EHA. 
Thus, it can be well understood that the difference
$B_{-\sigma} - B^{(1)}_{-\sigma}$ is contained in
(\ref{eq:iaalevels}) and (\ref{eq:iaacons}). If the
difference is neglected, we will get the atomic levels
and weights of the conventional AA again.

This concludes the theory. However, we still have to check whether 
the IAA self-energy is correct up to order $U^2$. This is by no
means obvious since the $U$ dependence of the atomic levels and 
the weights has become more complicated. Using the results 
(\ref{eq:iaalevels}) and (\ref{eq:iaacons}) in the CPA equation 
(\ref{eq:cpaeq}), we obtain for the IAA self-energy:
\begin{equation}
  \Sigma^{\rm {(IAA)}}_\sigma(E) = \frac{U n_{-\sigma}}{1 \; - \; 
  \frac{ \textstyle 
         \frac{1}{\hbar} \widetilde{G}_\sigma(E)
         \left( U - \Sigma^{\rm (IAA)}_\sigma(E) \right)
       }
       { \textstyle 
         1 \; - \; \frac{1}{\hbar} \widetilde{G}_\sigma(E)
         \left( B_{-\sigma} - B_{-\sigma}^{(1)} \right)
       }
  } \: .
\label{eq:sigfinal}
\end{equation}
The expression clearly shows the differences with respect to the 
conventional AA as well as to the EHA. The equation can be
expanded in powers of $U$ up to $U^2$ without knowing about the 
explicit $U$ dependence of the modified propagator. However, we 
need:
\begin{equation}
  B_{-\sigma} - B^{(1)}_{-\sigma} = \mbox{const.} \times U
  + {\cal O}(U^2) \: ,
\label{eq:bdiff}
\end{equation}
which can be verified easily \cite{PWN97}.
Using (\ref{eq:bdiff}), it turns out that up to order 
$U^2$ we have:
\begin{equation}
  \Sigma^{\rm {(IAA)}}_\sigma(E) = U n_{-\sigma}
  + U^2 n_{-\sigma} (1-n_{-\sigma}) \frac{1}{\hbar} 
  \widetilde{G}_\sigma(E) \: ,
\end{equation}
which is the same result as found in the EHA and thus proves the 
IAA to be exact up to order $U^2$ indeed (cf.\ Ref.\ \cite{WC94}):
\begin{equation}
  \Sigma^{\rm {(IAA)}}_\sigma(E) =
  U n_{-\sigma} + 
  \Sigma_{\sigma}^{\rm (SOC)}(E) + {\cal O}(U^3) \: .
\end{equation}

Let us briefly check the other limiting cases mentioned: It goes 
without saying that the trivial cases of $U=0$, $n=0$ or $n=2$ 
as well as the requirements imposed by particle-hole symmetry 
and the Herglotz properties \cite{MH73} are fulfilled. 
Suppressing the hopping of the $\downarrow$-electrons, we have 
$B_\uparrow = B_\uparrow^{(1)} = T_0$ from Eqs.\ (\ref{eq:bfin}) 
and (\ref{eq:b1def}). Hence, the theory reduces to the EHA and is 
thereby \cite{WC94} correct for the Falicov-Kimball model and also
in the atomic limit. By construction, the IAA self-energy has 
the correct high-energy expansion coefficients up to order $1/E^2$, 
and thus the moments of the spectral density $M_\sigma^{(0)}$ up 
to $M_\sigma^{(3)}$  are reproduced exactly. It can be shown 
analytically (similar to the discussion in Ref.\ \cite{PWN97})
or can simply be taken from the results presented in the next 
section that for strong $U$ the $\bf k$-resolved spectral 
density for each $\bf k$-point consists of two dominant peaks 
(Hubbard bands). Their energetic distance is roughly given by $U$. 
Since the first four moments of the spectral density are also 
conserved for each $\bf k$-point separately \cite{mkdisc}, the 
centers of gravity as well as the weights of both, the lower and 
the upper
Hubbard band, are fixed and can be calculated analytically for
$U\mapsto \infty$ along the lines of Ref.\ \cite{NB89}, for example.
The result turns out to be identical to what has been derived 
within the $t/U$-perturbation theory of Harris and Lange 
[see Eq.\ (\ref{eq:hl})].
Concludingly, the IAA improves upon the MAA with respect to the
weak-coupling and upon the EHA what concerns the strong-coupling 
regime. 
\\

{\center \bf \noindent VI. DISCUSSION AND RESULTS \\ \mbox{} \\} 

\mbox{}
\vspace{-7mm}

{\center \bf \noindent A. Paramagnetic phase \\ \mbox{} \\} 

For the symmetric case of half-filling and paramagnetism,
electron-hole symmetry requires $B_\sigma=B_\sigma^{(1)}=T_0$. 
This implies that the IAA reduces to the EHA in this case and thus 
cannot contribute
to an improved description of the Mott transition.
We therefore focus on the 
non-symmetric case $n\ne 1$. Fig.~1 shows
the spectral density of the $d=\infty$ Hubbard model on the 
hyper-cubic lattice for $U=4$ and $n=0.79$ as resulting
from the MAA, the EHA and the IAA (units are chosen such that
$t^\ast=1$). A non-zero temperature has been
taken ($\beta = 1 / k_{\rm B} T =7.2$) to enable a direct 
comparison with the essentially
exact QMC result of Jarrell and Pruschke \cite{JP93} which is 
also shown in the figure.

For each of 
the three approximate theories, the spectrum mainly consists of 
the two dominant charge-excitation peaks (Hubbard bands). 
Compared with the MAA
result, we notice that both peaks
are significantly broadened within the IAA, their spectral
weights being almost unchanged. 
Furthermore, while the MAA predicts nearly symmetric peak shapes, 
a clear peak asymmetry is
observed within the IAA spectrum.
The broadening effect can partly
be traced back to a stronger quasi-particle damping 
in the IAA compared with the MAA: The imaginary part of
the IAA self-energy is significantly 
larger at the energetic positions of the 
Hubbard bands.  

%+++++++++++++++++++++++++++++++++++++++++++++++++++++++++
\begin{figure}[t]
\vspace{-8mm}
\centerline{\psfig{figure=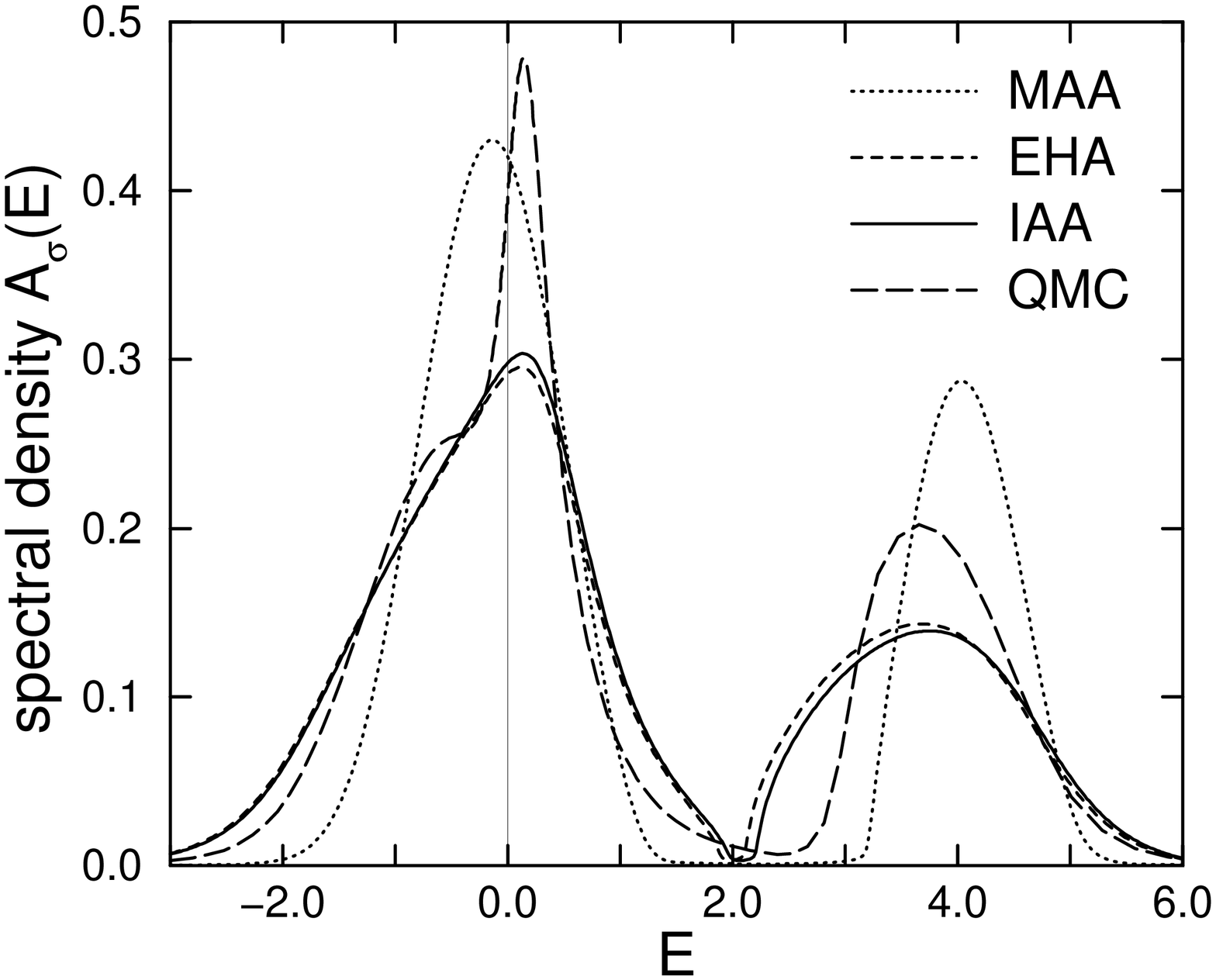,width=75mm,angle=270}}
\vspace{4mm}
\parbox[]{85mm}{\small Fig.~1.
Spectral density of the Hubbard model on the $d=\infty$ hyper-cubic
lattice for on-site Coulomb interaction $U=4$, band-filling 
$n=0.79$ and inverse
temperature $\beta = 7.2$. Results for the modified alloy-analogy
(MAA), the Edwards-Hertz approach (EHA) and the interpolating
alloy-analogy-based approximation (IAA) in comparison with 
quantum Monte Carlo data (QMC) from Jarrell and Pruschke 
\cite{JP93}.
All energies are given in units of $t^\ast =1$.
}
\end{figure}
%+++++++++++++++++++++++++++++++++++++++++++++++++++++++++

The IAA spectrum is found to be only slightly different from the
EHA result: Comparing with the EHA, we find the IAA spectral density
to be somewhat larger at its peak maximum in the lower Hubbard band.
There is more weight
in the lower and less weight in the upper peak which is 
shifted to higher energies compared with the EHA.

Besides the high-energy charge excitation peaks,
the spectrum obtained by QMC simulations \cite{JP93} shows
up a third structure around the Fermi edge at $E=0$, which is
reminiscent of the Kondo resonance. In the present situation, 
away from half-filling and for finite temperature, the resonance 
is strongly broadened but still clearly visible. On the other
hand, it is missing in the approximate approaches even for $T=0$
and increased band-filling, although the correct trend can be
recognized in the asymmetry of the IAA (and EHA) lower peak.
Compared with the QMC result, the MAA predicts a too large 
(quasi) gap between the peaks, while it is too small within the
IAA and EHA. The IAA (EHA) yields a width of the lower peak that
is quite close to the QMC result. The width of the upper peak,
however, comes out to large, while it is underestimated by the 
MAA. Passing from the EHA to the IAA, the agreement with the QMC 
result is improved very slightly.

%+++++++++++++++++++++++++++++++++++++++++++++++++++++++++
\begin{figure}[t]
\vspace{-8mm}
\centerline{\psfig{figure=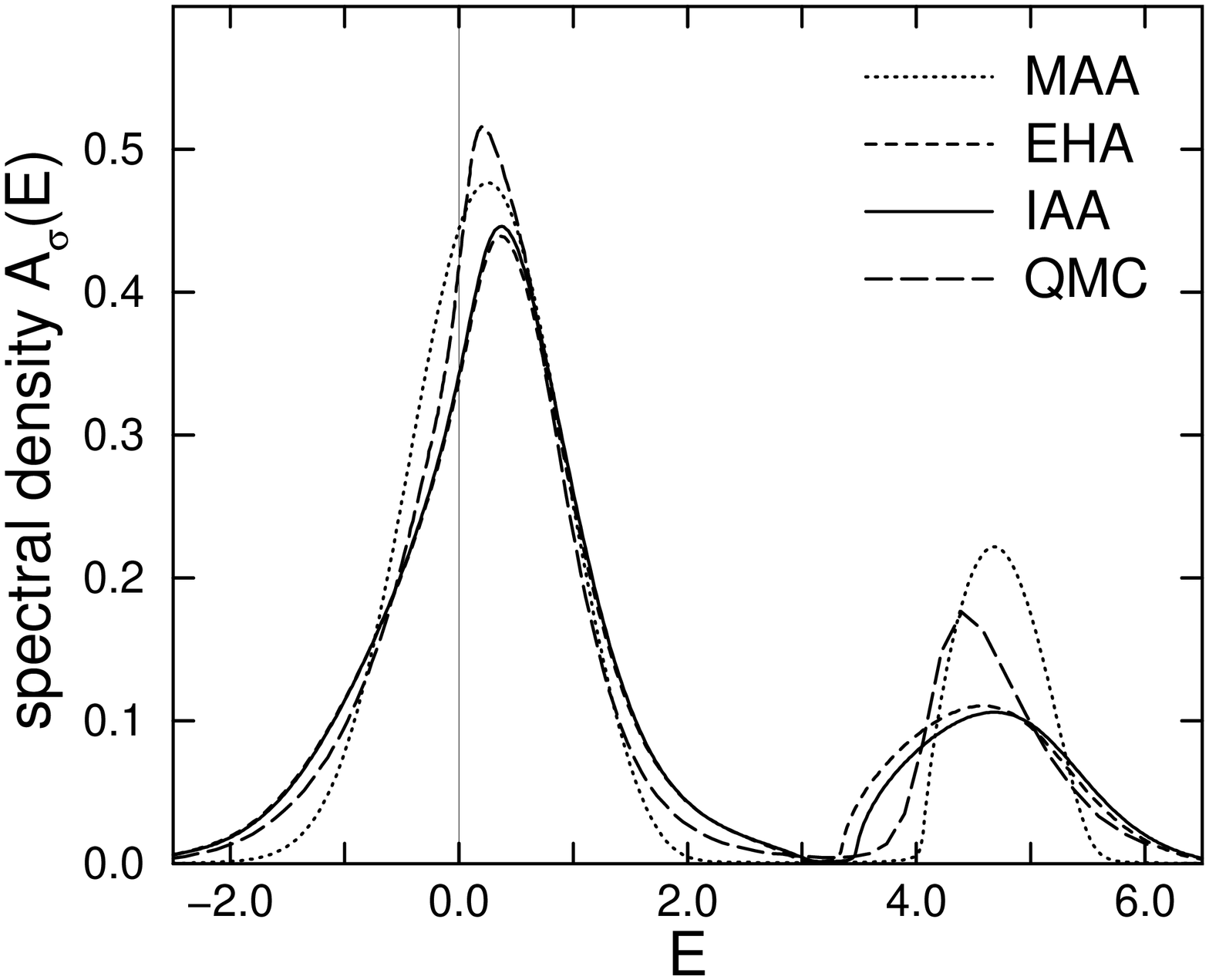,width=75mm,angle=270}}
\vspace{4mm}
\parbox[]{85mm}{\small Fig.~2.
The same as Fig.~1, but for smaller band-filling $n=0.57$.
}
\end{figure}
%+++++++++++++++++++++++++++++++++++++++++++++++++++++++++

%+++++++++++++++++++++++++++++++++++++++++++++++++++++++++
\begin{figure}[t]
\vspace{-8mm}
\centerline{\psfig{figure=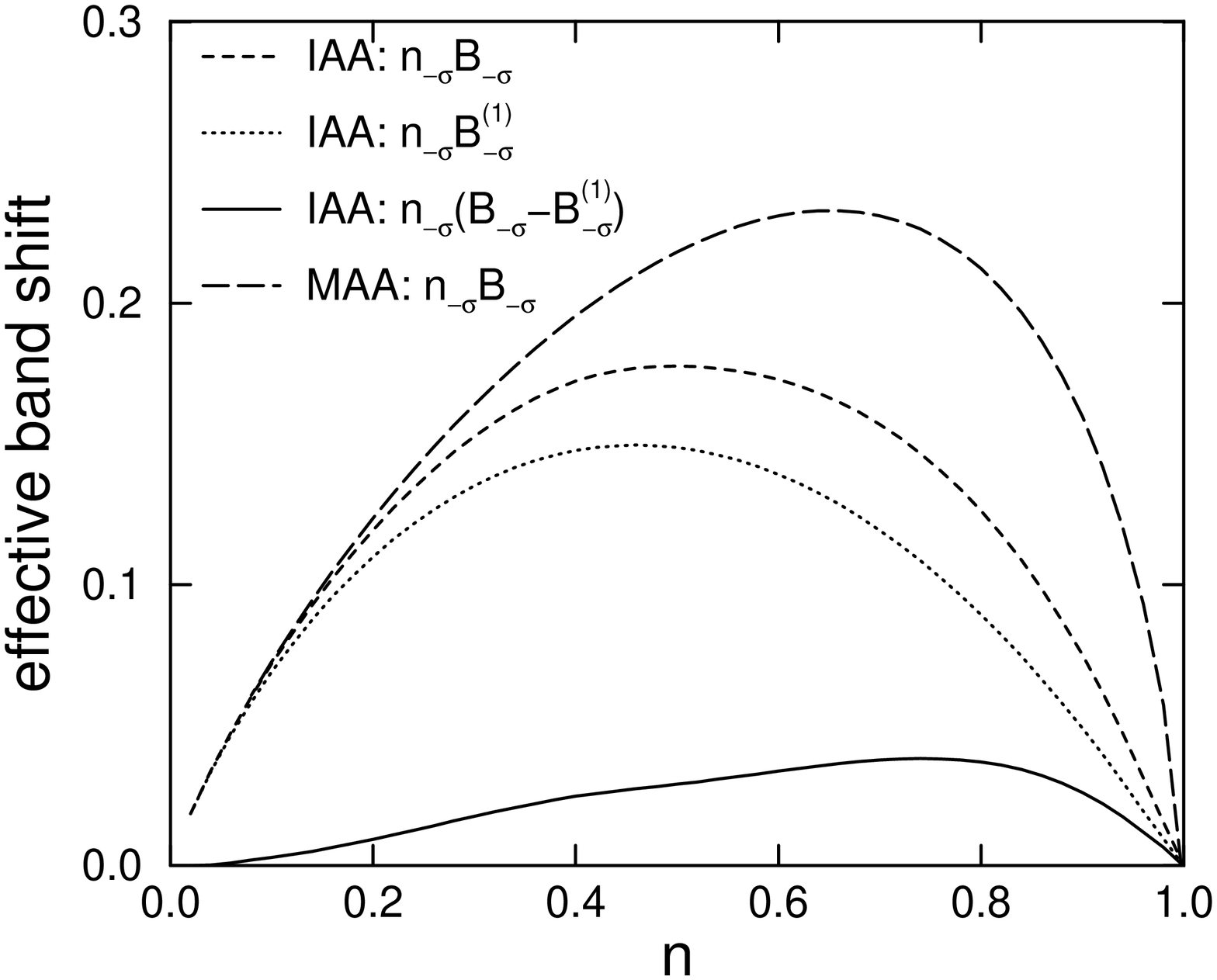,width=75mm,angle=270}}
\vspace{4mm}
\parbox[]{85mm}{\small Fig.~3.
Effective band shift $n_{-\sigma} B_{-\sigma}$ for the MAA and the
IAA and $n_{-\sigma} B^{(1)}_{-\sigma}$ (Hartree-Fock value, IAA)
as a function of band-filling $n$. Hyper-cubic lattice, $U=4$, 
$\beta=7.2$, $n_{\uparrow}=n_\downarrow$. 
Results are symmetric to the $n=1$ axis.
}
\end{figure}
%+++++++++++++++++++++++++++++++++++++++++++++++++++++++++

Decreasing the band filling, we find the same qualitative 
differences between the approximate approaches. However, they
become less pronounced. This can be
seen in Fig.~2 where the corresponding results are shown
for $n=0.57$. 
Comparing with the QMC spectrum, some discrepancies also
remain for lower $n$, but the overall agreement is certainly
improved for 
all approaches considered. This is mainly due to the fact that the 
Kondo-like resonance has disappeared in the QMC spectrum.

With respect to the EHA the main conceptual improvement of the IAA
consists in the corrected high-energy behavior of the self-energy.
This is determined by the correlation function $B_{-\sigma}$. 
In the strong-coupling regime $B_{-\sigma}$ leads to an 
effective shift of the lower Hubbard band 
$n_{-\sigma} B_{-\sigma}$ [cf.\  Eq.\ (\ref{eq:hl})]. This
is shown in Fig.~3 as a function of $n$. 
For the IAA as well as for the MAA the overall dependence on $n$ 
is quite similar to the results known from the spectral-density 
approach \cite{HN97b,HN96} as well as 
from the modified perturbation theory \cite{PWN97}.
We notice that (within the IAA) there is a small difference
between $B_{-\sigma}$ and its Hartree-Fock value 
$B_{-\sigma}^{(1)}$ only. Consequently, this implies via 
Eqs.\ (\ref{eq:sigeha}) and (\ref{eq:sigfinal}) that for 
the IAA only minor changes can be expected with respect to the EHA.
This is consistent with the results for the spectral density 
discussed above.
\\

{\center \bf \noindent B. Ferromagnetism \\ \mbox{} \\} 

Within the self-consistent calculation, the correlation 
functions $B_{-\sigma}$ might become spin-dependent. According
to Eq.\ (\ref{eq:hl}) this would
imply a spin-dependent shift of the centers of gravity of the 
two Hubbard bands. Therefore, the band shift $B_{-\sigma}$ is of 
exceptional importance for the possibility and thermodynamical 
stability of spontaneous ferromagnetic order.  

We consider two different cases where essentially
exact QMC results are available: the hyper-cubic
lattice with the BDOS given by Eq.\ (\ref{eq:bdoshc}), and
the BDOS (\ref{eq:bdosfcc}) corresponding to an fcc-type lattice.
A ferromagnetic instability is indicated by a divergency of the 
uniform static susceptibility which is calculated 
within the paramagnetic phase in the limit of an infinitesimally
small applied field $H$:
$\chi \sim \partial ( n_{\uparrow} - n_{\downarrow} ) / 
\partial H|_{H = 0}$.
Jarrell and Pruschke \cite{JP93} found a non-diverging
susceptibility within QMC on the hyper-cubic lattice for various 
fillings and temperatures and interactions up to $U=8$. This is
consistent with the findings of Refs.\ \cite{FMMH90,Uhr96} where it
is proved that the Nagaoka state \cite{Nag66} is instable for any 
$n\ne 1$. Let us mention, however, that a partially polarized 
ferromagnetic state has been found recently for $U\mapsto \infty$ 
within the non-crossing approximation \cite{OPK97}.

The QMC result for $U=4$ and $\beta=7.2$ shows $\chi$ to be a 
monotonically increasing function of the filling up to $n=1$ with a
slightly negative curvature, $\partial^2 \chi / \partial n^2 < 0$
\cite{JP93}. Fig.~4 shows the corresponding results from our 
analytical approaches. First, we notice a zero of $\chi^{-1}$ at 
$n=0.98$ within the IAA. A non-zero magnetization is found
for $0.98 \le n \le 1.02$ indeed. However, this has not been 
analyzed in more detail since obviously (comparing with QMC) 
it has to be interpreted as an artefact of the method. All three 
approaches considered cannot be expected to yield reliable results 
near to half-filling: In this range antiferromagnetic spin 
fluctuations are known to become important which manifest 
themselves in a narrow Kondo-like resonance at 
the Fermi level. The resonance, however, was shown to be absent 
in the analytical
approaches (see above). 

%+++++++++++++++++++++++++++++++++++++++++++++++++++++++++
\begin{figure}[t]
\vspace{-8mm}
\centerline{\psfig{figure=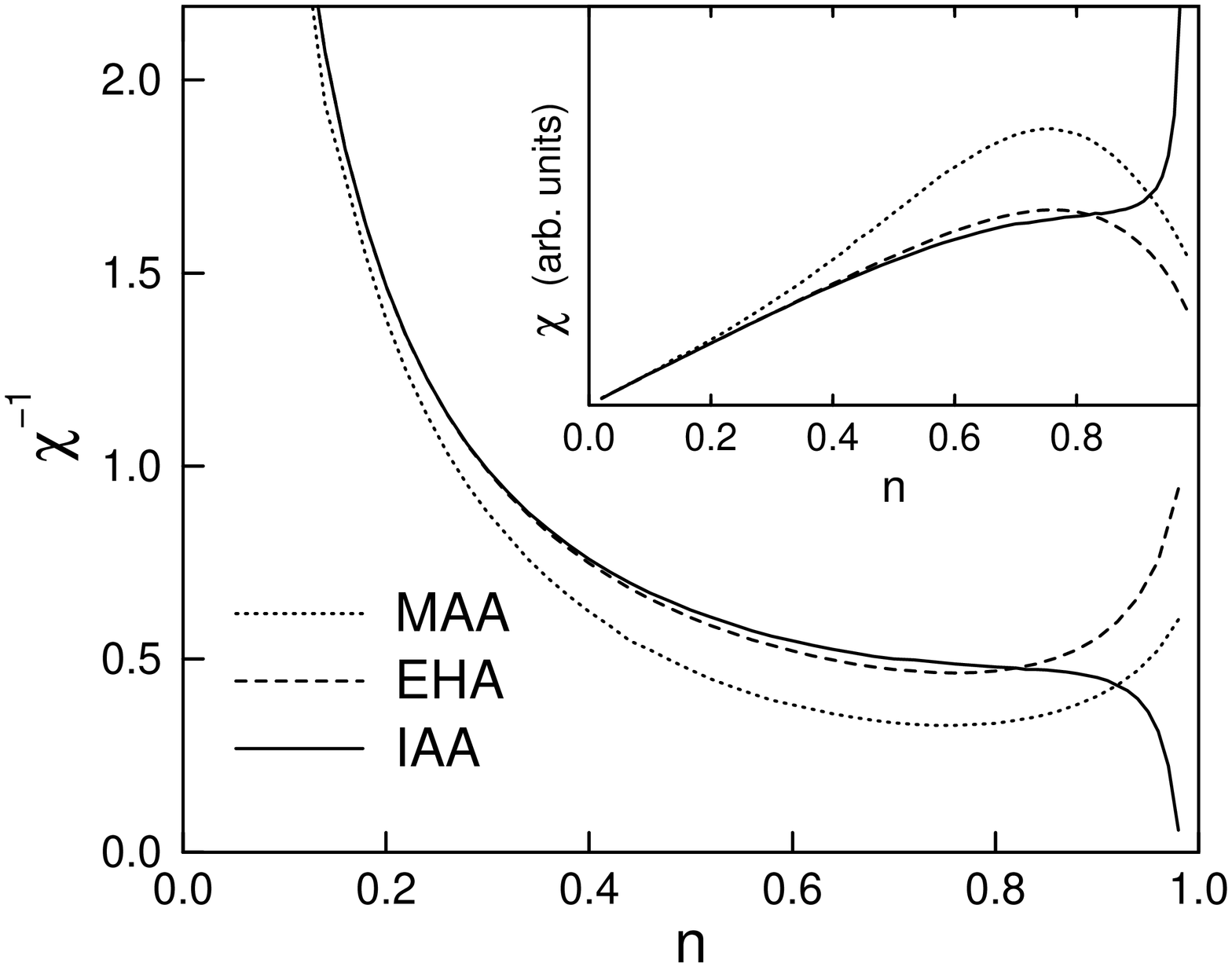,width=75mm,angle=270}}
\vspace{4mm}
\parbox[]{85mm}{\small Fig.~4.
Inverse static susceptibility $\chi^{-1}$ (inset: $\chi$) as a 
function of band-filling $n$ for the hyper-cubic
lattice, $U=4$ and $\beta=7.2$. Results for the MAA, EHA and IAA.
}
\end{figure}
%+++++++++++++++++++++++++++++++++++++++++++++++++++++++++

Well below half-filling the IAA predicts
a rather featureless positive susceptibility which increases with 
increasing $n$ and shows up a negative curvature. All this agrees 
with the QMC result. It should be mentioned that there are only 
minor changes in the results for temperature $T\mapsto0$. In 
particular, for the IAA we have checked that the artificial 
ferromagnetic solution is always confined to the very small range 
around half-filling.
The differences with respect to the EHA result are again found
to be small except for fillings close to $n=1$ where the results
are unphysical anyway. Finally, we notice that within 
the MAA the functional dependence of $\chi$ on $n$ shows up the
wrong curvature for smaller $n$
and disagrees in this respect with the QMC as 
well as with the IAA result.

A non-bipartite lattice as well as a strongly asymmetric 
BDOS are known to favor ferromagnetic order.
This has been demonstrated in a comprehensive way by recent
QMC studies \cite{VBH+97,Ulm98,WBS+97} and variational approaches
\cite{HUMH97}. The results of Uhrig \cite{Uhr96} and Ulmke 
\cite{Ulm98} show a ferromagnetic phase
to be stable in a fairly extended region of the
$U$-$n$-$T$ phase diagram for the $d=\infty$ fcc-type lattice. 
Ferromagnetism is favored due to the strong asymmetry of the BDOS
(\ref{eq:bdosfcc}) with its square-root divergency at the lower
band edge. 

%+++++++++++++++++++++++++++++++++++++++++++++++++++++++++
\begin{figure}[t]
\vspace{-8mm}
\centerline{\psfig{figure=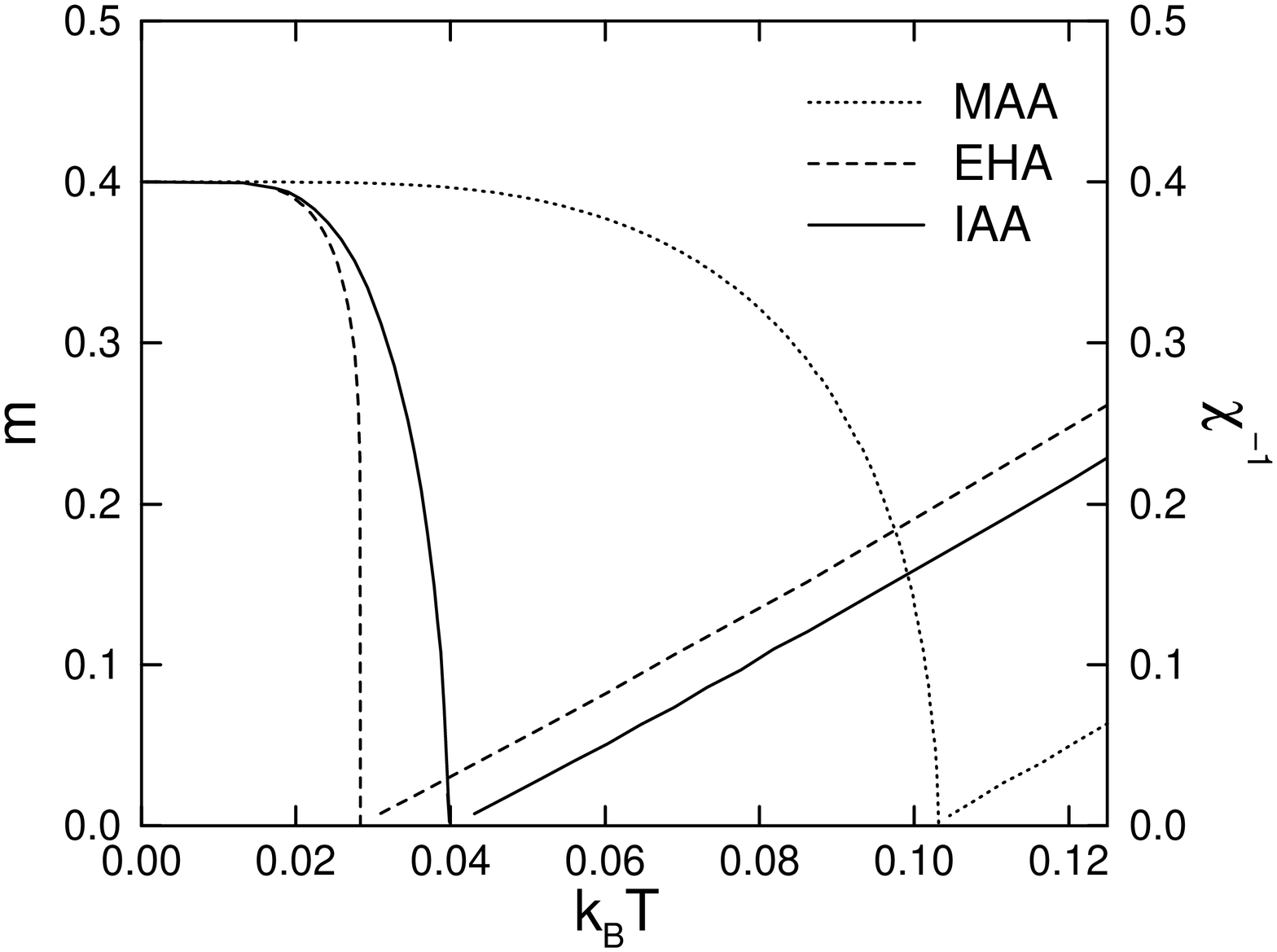,width=75mm,angle=270}}
\vspace{4mm}
\parbox[]{85mm}{\small Fig.~5.
Spontaneous magnetization 
$m=\langle n_{\uparrow} \rangle - \langle n_{\downarrow} \rangle$
and inverse static susceptibility $\chi^{-1}$
as functions of temperature for the Hubbard model on the 
$d=\infty$ fcc-type lattice [cf.\ BDOS in Eq.\ (\ref{eq:bdosfcc})]
for $U=4$ and $n=0.4$. Results for the MAA, EHA and IAA.
}
\end{figure}
%+++++++++++++++++++++++++++++++++++++++++++++++++++++++++

Thermodynamically stable ferromagnetic solutions are found 
within the MAA, the EHA as
well as in the IAA. Fig.~5 shows the temperature dependence of 
the spontaneous magnetization $m=n_\uparrow - n_\downarrow$ 
for $U=4$ ($t^\ast=1$) and 
$n=0.4$. In each case a fully polarized state ($m=n$) is obtained
for $T=0$. 
Increasing the temperature results in a decrease of $m$. The
magnetization curves are Brillouin-function-like, and the estimated
critical exponent of the order parameter is $\beta=1/2$, i.\ e.\
the expected mean-field value. The inverse susceptibility is also 
shown in Fig.~5. At the respective Curie temperature $T_C$ there 
is a zero of $\chi^{-1}$ as it must be for a second-order phase 
transition. For $T>T_C$ the temperature dependence of $\chi^{-1}$
is almost linear and thereby follows the Curie-Weiss law. 
For the critical exponent of the susceptibility we obtain the 
mean-field value $\gamma =1$.
The Curie temperature predicted by the IAA ($T_C=0.040$) almost
perfectly agrees with the essentially exact QMC result 
($T_C = 0.039$ \cite{Ulm98}).

For larger filling, at $n=0.58$, QMC data \cite{VBH+97,Ulm98} are 
available for the temperature dependence of $m$ and $\chi^{-1}$. 
The comparison with the MAA and IAA results is shown in Fig.~6. 
In both cases a stable ferromagnetic solution and a second-order 
transition to the paramagnetic phase is found.
For high temperatures the inverse static susceptibilities obey 
the Curie-Weiss law. As $T \mapsto T_C$, however, the
IAA inverse susceptibility deviates from the linear behavior. 
This is not observed for lower fillings (Fig.~5). The
zero of $\chi^{-1}$ determines the Curie temperature $T_C=0.031$
while the zero obtained from the extrapolation of the linear 
high-temperature trend yields the paramagnetic Curie temperature
$\Theta=0.057$. The correct (QMC) Curie temperature is found 
in between: $T_C=0.05$ \cite{VBH+97,Ulm98}; yet the agreement
with respect to $T_C$ is reasonable. However, the non-linear 
trend of $\chi^{-1}(T)$ near $T_C$ is at variance with the QMC
result. Also the $T=0$ magnetization ($m=0.47$) comes out too 
small since the Brillouin-function fit to the QMC data for $m$ 
indicates a fully polarized ground state ($m=n=0.58$). 

%+++++++++++++++++++++++++++++++++++++++++++++++++++++++++
\begin{figure}[t]
\vspace{-8mm}
\centerline{\psfig{figure=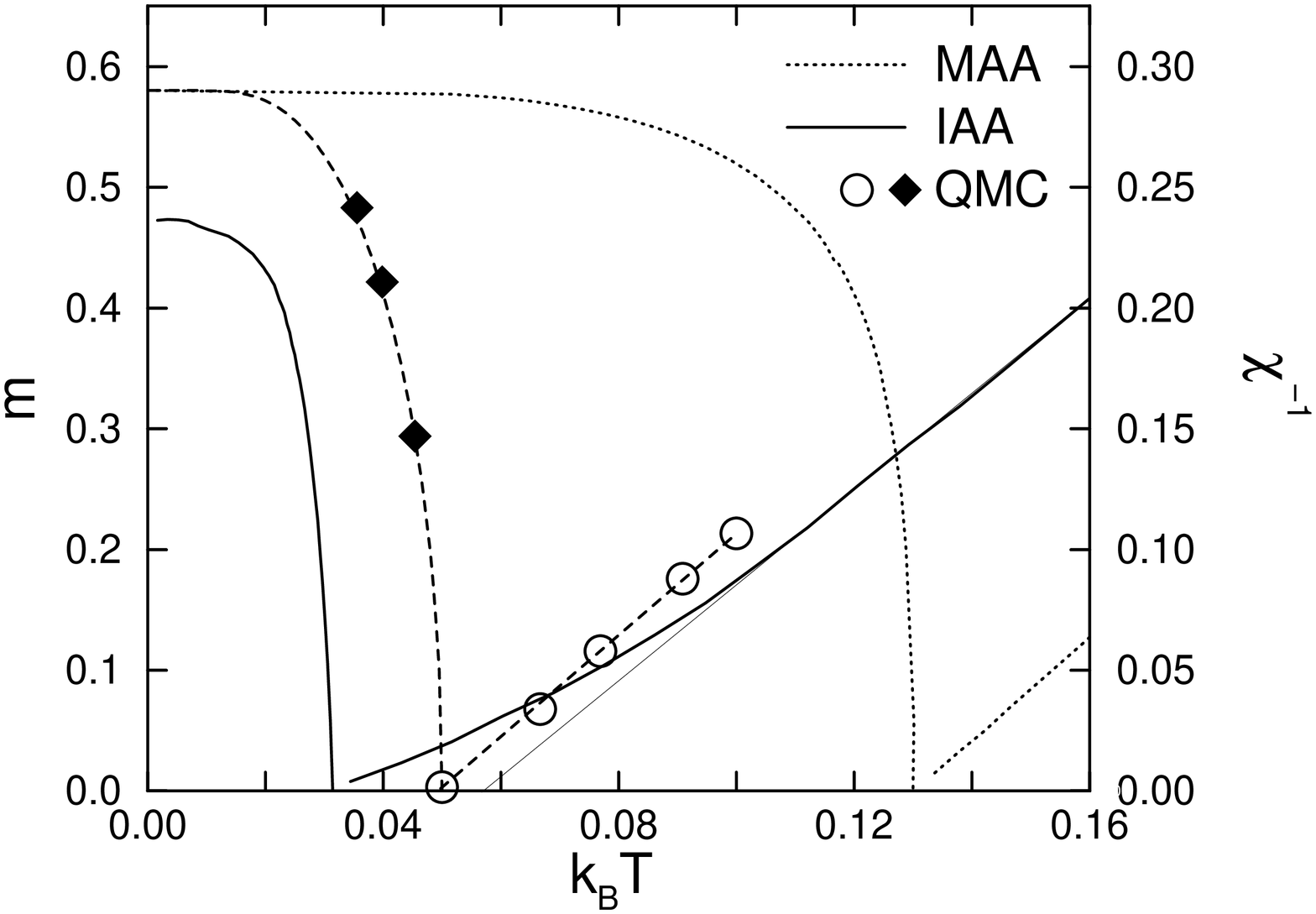,width=68mm,angle=270}}
\vspace{4mm}
\parbox[]{85mm}{\small Fig.~6.
Temperature dependence of $m$ and $\chi^{-1}$ for $U=4$ and 
n=0.58. Results for the MAA and IAA in comparison with the 
quantum Monto Carlo data (QMC) from Vollhardt et al. \cite{VBH+97}
(diamonds/circles, dashed line: linear fit to $\chi^{-1}$ / fit 
with a Brillouin function to $m$). The thin solid line is a
linear extrapolation of (the IAA) $\chi^{-1}(T)$ to lower 
temperatures.
}
\end{figure}
%+++++++++++++++++++++++++++++++++++++++++++++++++++++++++

The IAA results for $n=0.58$ appear to be less reliable than
those for $n=0.4$. Still, the IAA turns out to be superior 
compared to the MAA, which yields a considerably too high Curie 
temperature ($T_C=0.130$), and also to the EHA:
At $n=0.58$ the EHA does not yield a ferromagnetic solution at all,
and we are left with the paramagnetic phase only. 

%+++++++++++++++++++++++++++++++++++++++++++++++++++++++++
\begin{figure}[t]
\vspace{-8mm}
\centerline{\psfig{figure=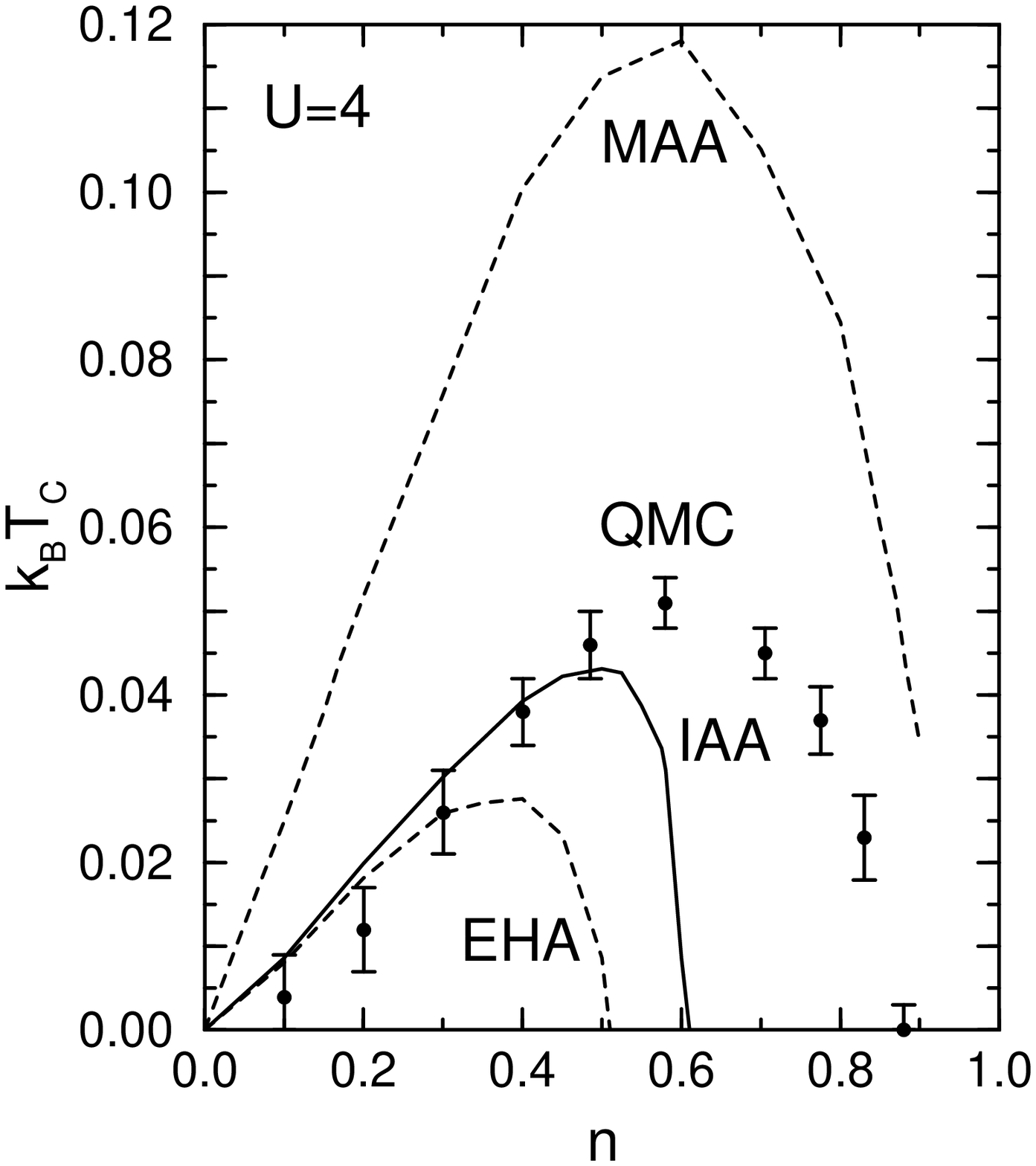,width=95mm,angle=0}}
\vspace{-25mm}
\parbox[]{85mm}{\small Fig.~7.
Filling dependence of the Curie temperature at $U=4$ within the MAA,
EHA and IAA
in comparison with the QMC result (error bars) by Ulmke \cite{Ulm98}.
}
\end{figure}
%+++++++++++++++++++++++++++++++++++++++++++++++++++++++++

%+++++++++++++++++++++++++++++++++++++++++++++++++++++++++
\begin{figure}[t]
\vspace{4mm}
\centerline{\psfig{figure=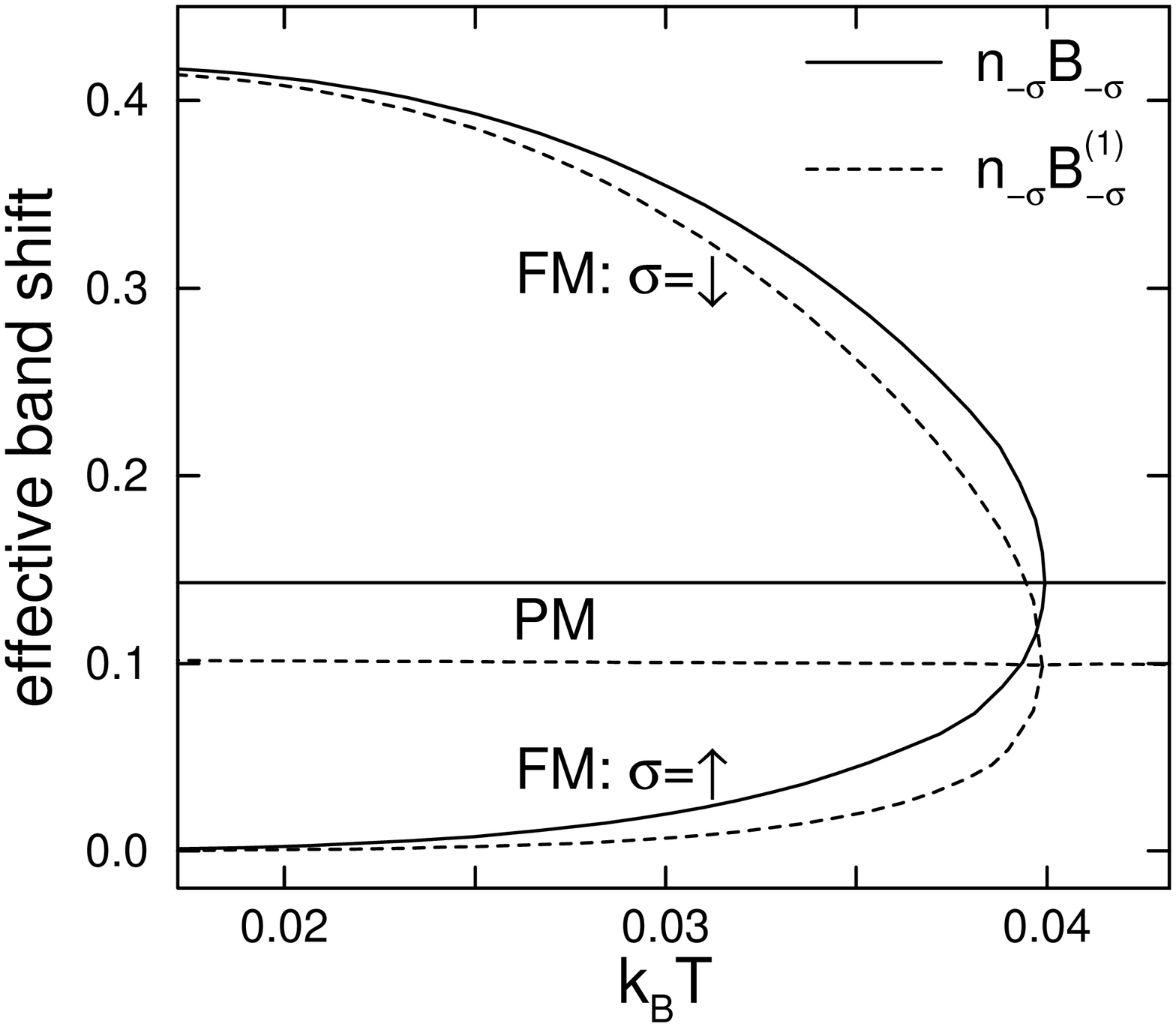,width=70mm,angle=270}}
\vspace{4mm}
\parbox[]{85mm}{\small Fig.~8.
Effective band shift $n_{-\sigma} B_{-\sigma}$ and Hartree-Fock
value $n_{-\sigma} B^{(1)}_{-\sigma}$ calculated within the IAA
as a function of temperature for the ferromagnetic (FM) and the
paramagnetic phase (PM). $d=\infty$ fcc lattice, $U=4$, $n=0.4$.
}
\end{figure}
%+++++++++++++++++++++++++++++++++++++++++++++++++++++++++

%+++++++++++++++++++++++++++++++++++++++++++++++++++++++++
\begin{figure}[t]
\vspace{-10mm}
\centerline{\psfig{figure=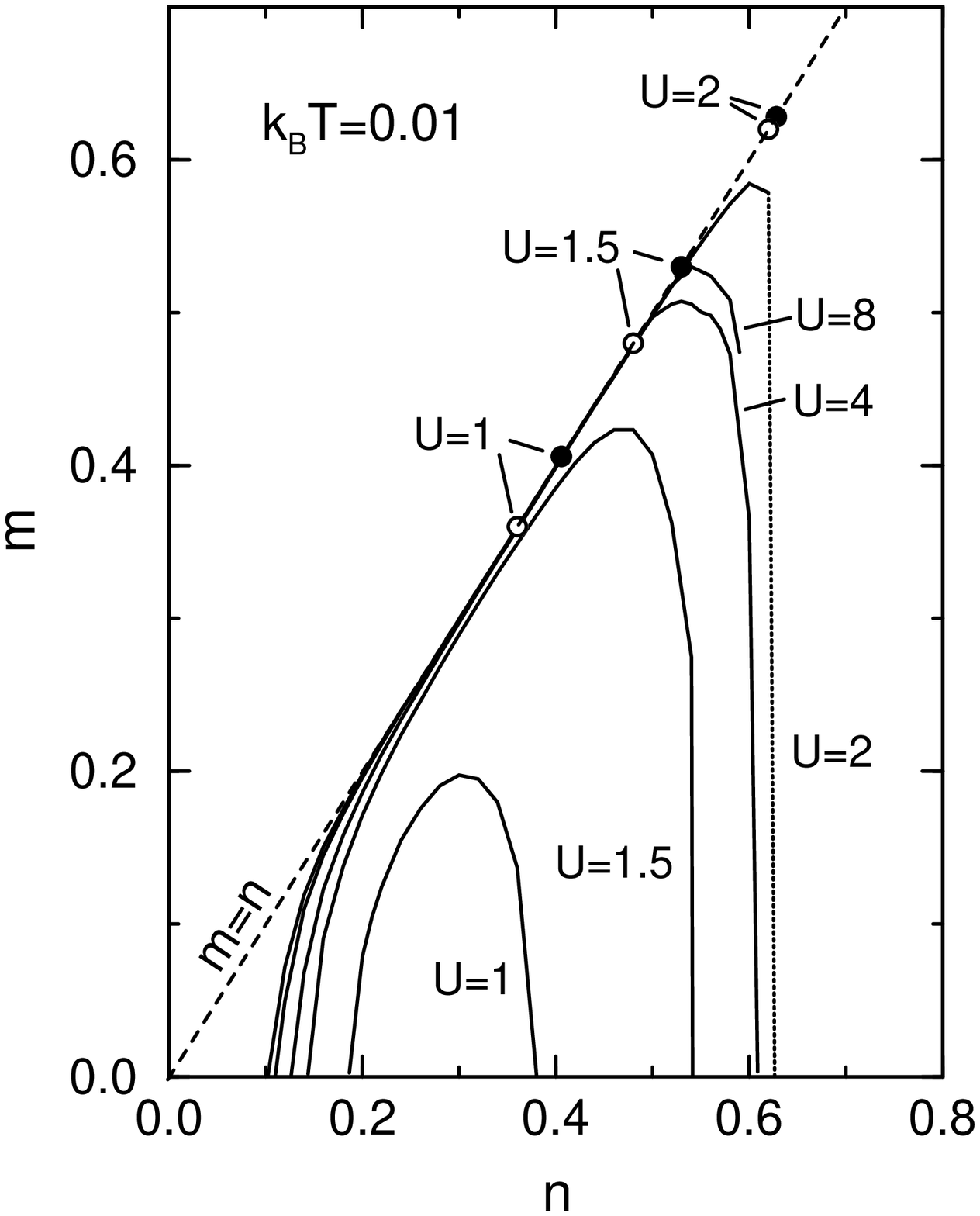,width=100mm,angle=0}}
\vspace{-17mm}
\parbox[]{85mm}{\small Fig.~9.
Filling dependence of the magnetization at $k_BT=0.01$ and different
values of $U$. The lower critical filling $n_{c,1}(U)$, above
which ferromagnetism occurs for $k_BT=0.01$, decreases with 
increasing $U$. The open circles indicate the filling $n_N(U)$
below which the system remains fully polarized at $T=0$:
$n_N(1.0)=0.36\pm 0.03$, $n_N(1.5)=0.48\pm 0.02$, 
$n_N(2.0)=0.62\pm 0.02$
(errors due to the extrapolation to $T=0$). For comparison the
exact results for $n_N$ by Uhrig \cite{Uhr96} are also shown
(filled circles).
}
\end{figure}
%+++++++++++++++++++++++++++++++++++++++++++++++++++++++++

In the discussion of the paramagnet, it has been mentioned that 
because of the absence of the Kondo-like feature in the IAA spectrum, 
reliable results cannot be expected 
for fillings close to half-filling. 
A restricted range of validity is also found with respect to the 
magnetic properties. The IAA is able to predict a reliable value for 
the Curie temperature for fillings up to $n\approx 0.5$ only. Below
$n=0.5$, however, the agreement between the IAA and QMC is convincing:
Fig.~7 shows the calculated filling dependence of $T_C$ for $U=4$.
The phase transition is always of second order. 
For $n<0.5$ the IAA result falls into the error bounds of the QMC 
data (except for $n=0.2$). Possibly, there is a slight tendency to 
overestimate $T_C$ in general. For $n>0.5$ the Curie temperature 
steeply decreases with increasing filling. Beyond $n>0.61$ only 
the paramagnetic solution can be found, while the ferromagnetic 
region extends up to $n\approx 0.88$ within QMC.

The MAA considerably overestimates the possibility 
for ferromagnetism.
Over the whole $n$ range considered, the Curie temperature is too
large by more than a factor 2. Ferromagnetic solutions exist for
$0 < n \lesssim 1$ (close to half-filling we encountered numerical
difficulties to stabilize truly converged solutions). The $T_C(n)$
obtained by means of the EHA is similar to the IAA curve. As can be
seen in Fig.~7, however, $T_C$ is too small compared with the QMC
result if $n > 0.3$, and the ferromagnetic region shrinks to 
$n<0.51$ in the EHA, thereby underestimating the possibility for
ferromagnetic order. The IAA, being a combination of the MAA and 
the EHA, yields a $T_C$ in between and the best agreement with the
QMC result.

The interpretation for the different transition temperatures in the
different approaches is the following: That there is a too high $T_C$
within the MAA can be attributed to the fact that quasi-particle 
damping is underestimated compared with the IAA (cf.\ preceding 
subsection). Damping effects tend to reduce the density of states 
at the Fermi edge which according to Stoner's criterion implies a 
lower $T_C$ \cite{HN96}. The difference in the Curie 
temperatures between the EHA and IAA must 
result from the approximation
$B_{-\sigma} \mapsto B_{-\sigma}^{(1)}$. The temperature dependence
of the corresponding effective band shifts is shown in Fig.~8 for 
$n=0.4$. In the paramagnetic phase 
$n_{-\sigma} B_{-\sigma}$ and $n_{-\sigma} B^{(1)}_{-\sigma}$ 
are (almost) temperature independent. In the ferromagnetic phase
below $T_C$ there is a spin splitting of both quantities. As 
$T\mapsto 0$ the fully polarized state is reached, and 
$n_{\downarrow} = 0$ implies a vanishing effective band shift
$n_{\downarrow} B_{\downarrow} = n_{\downarrow} B^{(1)}_{\downarrow}
= 0$. On the other hand, for $T\mapsto 0$
we have $\Sigma_{\uparrow}(E) \mapsto 0$ which implies via Eqs.\
(\ref{eq:bfin}) and (\ref{eq:b1def}) that 
$n_{\uparrow} B_{\uparrow} - n_{\uparrow} B^{(1)}_{\uparrow}
\mapsto 0$. Fig.~8 shows that in both cases, 
$\sigma = \uparrow, \downarrow$, the difference
$n_{-\sigma} B_{-\sigma} - n_{-\sigma} B^{(1)}_{-\sigma}$
increases with increasing $T$ and is at its maximum for $T=T_C$,
i.~e.\ for the paramagnet. For the ferromagnet we can conclude
that replacing $B_{-\sigma}$
by $B_{-\sigma}^{(1)}$ in the {\em self-consistent} solution for
the self-energy, will result in small changes of the spin-dependent
spectral density only. These are comparable to those discussed 
in the previous 
subsection for the paramagnet. More important, however, {\em within}
the self-consistency cycle the effective band shift 
can have a strong influence on the magnetic aspects of the final
solution. 
Therefore, using the Hartree-Fock approximation for $B_{-\sigma}$ 
may be insufficient to describe the magnetic 
properties of the system. 

Analyzing the continued-fraction expansion of the Green function for
$T=0$, Uhrig \cite{Uhr96} could obtain the exact boundary $U=U_N(n)$ 
above which the fully polarized state is stable against a single spin 
flip for the fcc-type lattice. Due to the divergence of the BDOS at 
the lower band edge, the Nagaoka state is locally stable even for low 
fillings $n \mapsto 0$, the (in-)stability line smoothly ends at 
$U_N(0)=0$. To compare 
with these results, Fig.~9 shows the filling-dependence of the 
magnetization for different $U$ at a low temperature $T=0.01$.
At low fillings the system is paramagnetic. Above a critical
filling $n_{c,1}(U)$ a non-zero magnetization is found which steeply
increases with increasing $n$ and (almost) reaches the saturation
value $m=n$. Finally, for higher $n>n_{c,2}(U)$ ferromagnetism 
breaks down again. The higher critical filling $n_{c,2}(U)$ exhibits
an unusual $U$ dependence: It is at its maximum for $U=2$ 
and decreases
for larger $U>2$. This is believed to be an artefact of the method; 
we expect the IAA to be able to yield reliable information up to
$n \approx 0.5$ only. On the other hand, the $U$ dependence of 
the lower critical filling is plausible: $n_{c,1}(U)$ monotonically
decreases with increasing $U$. We have inspected the IAA 
result for $U=1$ in more detail for lower temperatures. Even at
$T=0.001$ the system is not fully polarized for low fillings. We 
found it difficult to access still lower temperatures numerically.
Extrapolating the temperature dependence of the magnetization 
to $T=0$, however, we find our results to be consistent with a
ferromagnetic phase extending to $n=0^+$. For larger fillings
extrapolation to $T=0$ yields the value $n_N$ up to 
which the system remains fully polarized. Fig.~9 shows that the
obtained fillings $n_N$ for $U=1.0, 1.5, 2.0$ are close to the 
results found by Uhrig \cite{Uhr96}. For the larger 
values of $U$ ($U > 2$) the IAA 
is not able reproduce $n_N(U)$, e.~g.\ a partially polarized ground 
state for $U=4$ and $n=0.58$ is inconsistent with the results of 
Ref.\ \cite{Uhr96} but is found in the IAA (see Fig.~6).

%+++++++++++++++++++++++++++++++++++++++++++++++++++++++++
\begin{figure}[t]
\vspace{-8mm}
\centerline{\psfig{figure=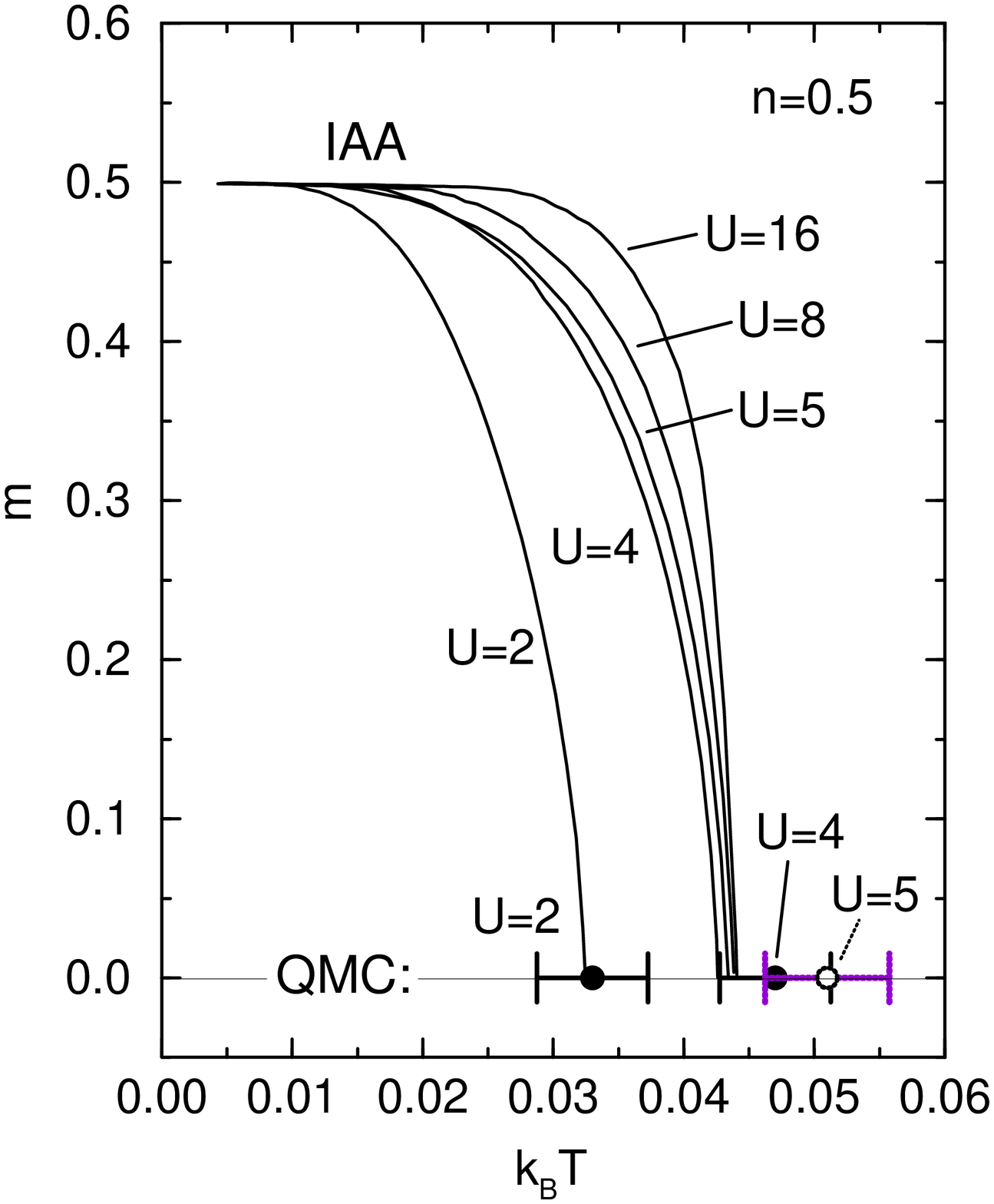,width=95mm,angle=0}}
\vspace{-8mm}
\parbox[]{85mm}{\small Fig.~10.
Temperature dependence of the magnetization at $n=0.5$ and different 
values of $U$. The QMC data for $T_C$ at $U=2, 4, 5$ are taken from 
the interpolation lines in Fig.~2 of Ref.\ \cite{Ulm98}; the bars 
indicate the typical (QMC) errors.
}
\end{figure}
%+++++++++++++++++++++++++++++++++++++++++++++++++++++++++

%+++++++++++++++++++++++++++++++++++++++++++++++++++++++++
\begin{figure}[t]
\vspace{4mm}
\centerline{\psfig{figure=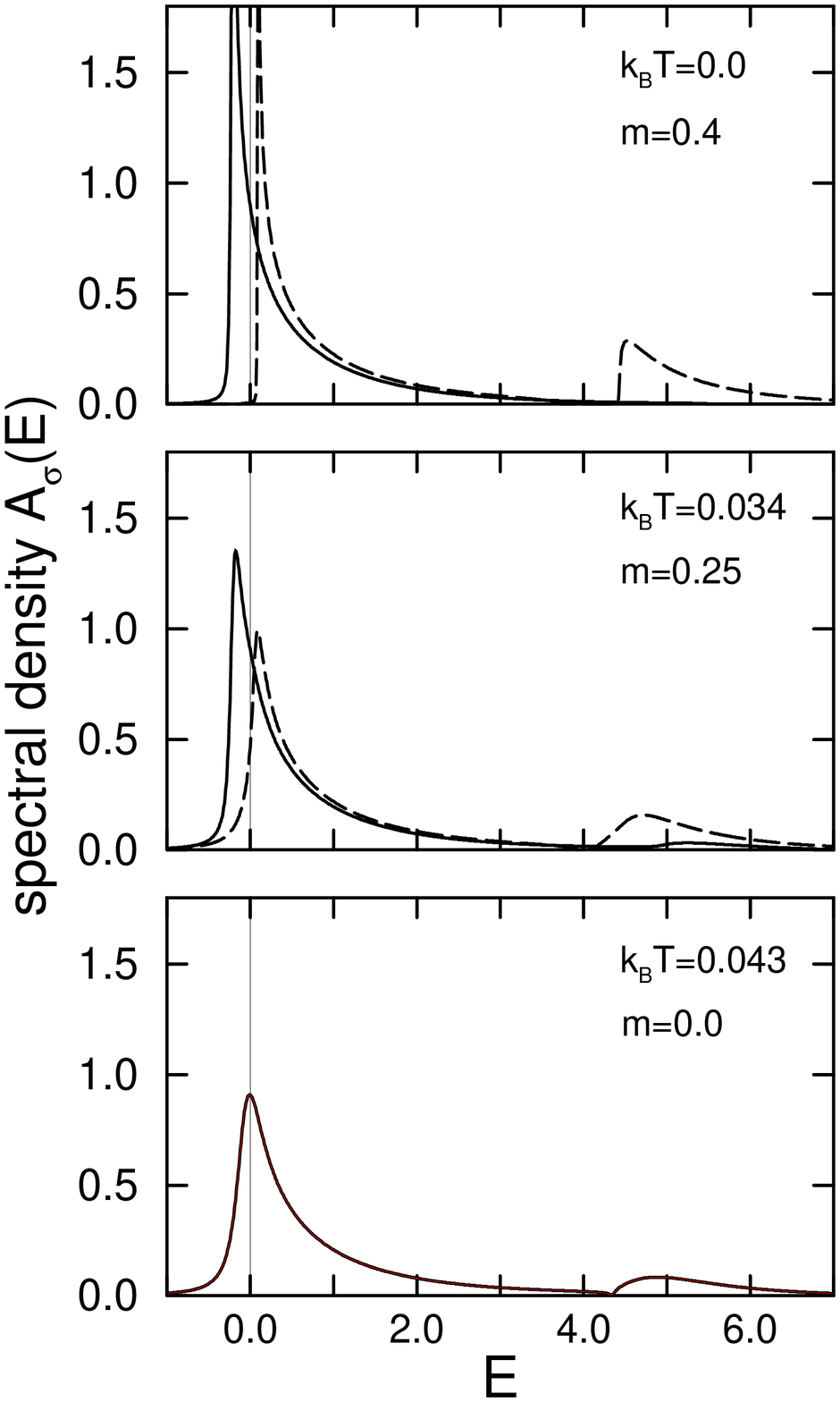,width=75mm,angle=0}}
\vspace{4mm}
\parbox[]{85mm}{\small Fig.~11.
Spin-dependent spectral density in the ferromagnetic phase for 
three different temperatures ($d=\infty$ fcc lattice, $U=4$, 
$n=0.4$). Solid line: majority spin ($\uparrow$). Dashed line: 
minority spin ($\downarrow$). 
$k_B T=0.043$ corresponds to the Curie temperature.
}
\end{figure}
%+++++++++++++++++++++++++++++++++++++++++++++++++++++++++

%+++++++++++++++++++++++++++++++++++++++++++++++++++++++++
\begin{figure}[t]
\vspace{-30mm}
\centerline{\psfig{figure=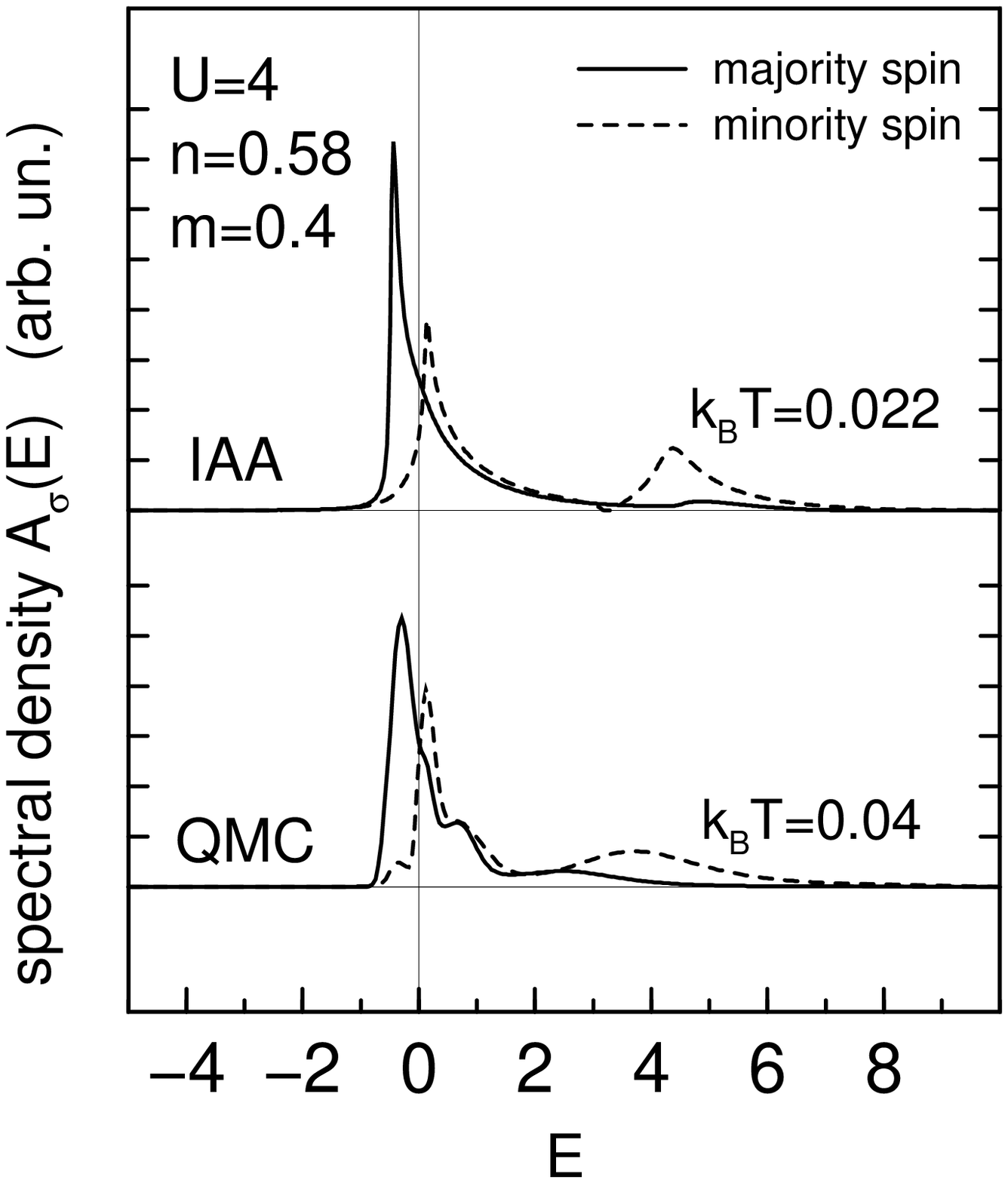,width=95mm,angle=0}}
\vspace{-12mm}
\parbox[]{85mm}{\small Fig.~12.
Spin-dependent spectral density for $U=4$ and $n=0.58$. 
Comparison between the IAA result for $k_BT=0.022$ and the QMC
result from \cite{Ulm98} for $k_BT=0.04$. In both cases
the magnetization is $m=0.4$.
}
\end{figure}
%+++++++++++++++++++++++++++++++++++++++++++++++++++++++++

To estimate to which strength of the interaction $U$ the IAA is 
a reliable approach, we have calculated the temperature-dependent
magnetization for different $U$ at the filling $n=0.5$. The results
are shown in Fig.~10. 
For each $U$ considered we observe ferromagnetic 
saturation for $T\mapsto 0$ and a second-order transition to the 
paramagnetic phase at the respective Curie temperature $T_C(U)$. 
As a function of $U$ the Curie temperature is found to saturate with 
$T_C(U=\infty) \approx 0.044$. We also included the predictions
of the QMC method for $T_C(U)$ in Fig.~10. The data for $U=2, 4, 5$
are taken from the values of the interpolation lines at $n=0.5$ in
Fig.~2 of Ref.\ \cite{Ulm98}. The error bars represent the respective 
{\em typical} error as has been estimated from the real data points 
and errors in the vicinity of $n=0.5$. There is a 
convincing agreement between IAA and QMC for $U=2$.  At $U=4$
the IAA result for $T_C$ still falls into the error bounds, while 
a slightly too low $T_C$ is predicted for $U=5$. We conclude that 
for strong interaction $U$ the IAA appears to be reasonable but 
less convincing compared with the low and intermediate $U$ regime
\cite{nca}. 

The temperature-dependence of the spin-resolved spectral densities
is shown in Fig.~11. We have taken $n=0.4$ and $U=4$ where the IAA
should yield reliable results. For $T=0$ the shape of the $\uparrow$ 
spectral density is given by the shape of the free BDOS 
(\ref{eq:bdosfcc}) since the magnetization is saturated, $m=n$, 
and thus $\Sigma_\uparrow \equiv0$. The square-root singularity at 
the lower band edge has been slightly smeared 
out for numerical reasons. We have checked that this does not 
affect the presented results for $m$, $\chi^{-1}$ and $T_C$.
The $\downarrow$ spectral density is split into the lower and
the upper Hubbard band, the energetic distance of which is roughly 
given by $U$. For $U=4$ the result of Harris and
Lange should apply. The energetic shift between the $\uparrow$ band
and the lower $\downarrow$ Hubbard band (centers of gravity) is 
read off from Fig.~11 to be 
$T_{1\downarrow} - T_{1\uparrow} \approx 0.40$.
On the other hand, the calculated effective band shift for the 
$\downarrow$ band is $n_\uparrow B_\uparrow = 0.42$ (see also 
Fig.~8). This is consistent with Eq.\ (\ref{eq:hl}) which for the 
present case and averaged over $\bf k$ simplifies to 
$T_{1\downarrow} - T_{1\uparrow} = n_\uparrow B_\uparrow$.

With increasing temperature the upper Hubbard band in the 
$\uparrow$ spectral density arises. There is an exchange splitting 
between the $\uparrow$ and $\downarrow$ upper Hubbard bands being 
inverse to the splitting between their lower counterparts. For 
$T\mapsto T_C$ the spin splitting 
becomes gradually smaller. At the same time we observe a transfer
of spectral weight between the two spin channels that speeds up the
depolarization of the system. At $T=T_C$ both spectra have become
identical. Compared with the $T=0$ case, the singularity at the 
lower band edge has disappeared, and a strong broadening of the
spectrum can be observed which is due to quasi-particle damping.

The qualitative behavior of the spectral density as a function of
temperature, i.~e.\ what concerns the spin-dependent centers of 
gravity, effective widths and spectral weights of the Hubbard bands,
very well agrees with the results of Harris and Lange. This 
implies that the higher-order corrections in the $t/U$ expansion
that have been neglected in Eq.\ (\ref{eq:hl}) can be considered
to be small (for $U=4$). 

At the same interaction strength, but for higher filling there is 
one QMC result available for the spin-dependent spectral density 
in the ferromagnetic phase \cite{Ulm98}. Unfortunately, the filling 
($n=0.58$) and the temperature ($T=0.04$) where the QMC spectrum 
has been obtained, are not well suited for a one-to-one comparison 
between IAA and QMC since the Curie temperature predicted by the 
IAA is below $T=0.04$ (see Fig.~6). We therefore try a comparison 
of the QMC spectrum with the IAA result at $T=0.022$ where the 
magnetization is non-zero and equal to the magnetization found 
within QMC ($m=0.4$). Fig.~12 shows the resulting spin-dependent 
spectral densities. A quantitative agreement with the QMC spectrum 
cannot be expected: Due to the temperature dependence of the 
effective band shift $n_{-\sigma} B_{-\sigma}$, the positions and 
weights of the Hubbard bands cannot be the same as those in QMC 
spectrum at a different temperature. Qualitatively, we notice that
apart from the Hubbard bands there is an additional peak in the 
$\uparrow$ and the $\downarrow$ QMC spectrum at $E\approx 0.8$ and
also a peak at $E\approx -0.4$ in the $\downarrow$ channel
which is not seen in the IAA result. At zero energy we observe a 
weak shoulder in the $\uparrow$ QMC spectrum reminiscent of the 
Kondo resonance. A (much weaker) 
structure at $E=0$ is also found by the IAA.
\\

{\center \bf \noindent VII. CONCLUSION \\ \mbox{} \\}

Exactly solvable limiting cases represent strong necessary 
conditions for any approximate solution of the
Hubbard model. The central idea of the present paper has been to
develop an analytical expression for the self-energy that is 
compatible with as much rigorous results as possible. From this
point of view, the interpolating alloy-analogy-based approximation 
(IAA) can be considered as one of the best analytical approaches
that are available for the Hubbard model in infinite dimensions. 

It avoids the defects of the modified alloy analogy (MAA)
\cite{HN96} which
is not able to recover Fermi-liquid properties 
in the weak-coupling regime, and it improves upon the
Edwards-Hertz approximation (EHA)
\cite{EH90,WC94} which is at variance with the
exact strong-coupling results of Harris and Lange. The IAA not
only recovers the trivial $U=0$ and the $t=0$ limits but also the 
respective first non-trivial corrections: For small $U$ it is exact 
up to order $U^2$ and can thus be regarded as a generalization of 
standard weak-coupling perturbation theory. For strong $U$ it yields
the correct positions and weights of the Hubbard bands which are 
exactly known from perturbation theory in $t/U$. Formally, the IAA 
is constructed by a proper combination of the MAA with the EHA: The 
general form of the ansatz for the self-energy it taken from the 
EHA, but following the concept of the MAA, the ``atomic levels'' 
and ``concentrations'' are determined via comparison with the 
first four exactly known moments of the spectral density. As a 
consequence, the coefficients of the high-energy expansion of 
the self-energy are exact up to order $1/E^2$. Furthermore, when 
applied to the Falicov-Kimball model, i.~e.\ when suppressing the 
hopping of one spin species, the IAA reduces to the conventional 
alloy-analogy solution which has been shown to be exact in 
$d=\infty$ by Brandt and Mielsch.

The main disadvantage of the method consists in the fact that it
cannot be motivated from a simple and physical concept. 
Inherently, its justification rests on its interpolating 
character only. In particular, the IAA is not conserving in 
the sense of Baym and Kadanoff. This must be taken into account
when interpreting Fermi-surface and related properties. For the 
symmetric case of a paramagnet at half-filling the IAA simply 
reduces to the EHA which predicts the Mott transition but also a 
non-Fermi-liquid metallic phase. Within the EHA and similarly for 
the IAA, the latter turns out to be existing in a 
fairly extended region of the $T=0$ phase diagram. On the other 
hand for $d=\infty$ there
are no indications for non-Fermi-liquid behavior in combined
studies with iterative perturbation theory and quantum
Monte Carlo (QMC) simulations \cite{GKKR96}. 
The non-Fermi-liquid metallic phase found in the IAA must be tightly
connected with the alloy-analogy basis of the approximation and
thus with the limit of the Falicov-Kimball model since it is known
that in this limit the (mobile) electrons indeed do not form a 
Fermi liquid \cite{Vol93}.

For an a posteriori test of the quality of the IAA, we have 
compared the results with numerically exact quantum Monte Carlo 
data from Jarrell and Pruschke \cite{JP93} as well as from Ulmke
and Vollhardt at al.\ \cite{VBH+97,Ulm98}, with the results of 
Uhrig concerning the (in-)stability of the Nagaoka state \cite{Uhr96} 
and with the results of the MAA and the EHA, mainly for the decisive 
moderate- to strong-coupling regime.

For the paramagnet on the hyper-cubic lattice, the differences 
between the EHA and IAA results are rather small, and judged from
the QMC data, the IAA only slightly improves upon the MAA and EHA. 
At fillings close to half-filling the IAA fails to reproduce the
Kondo-like quasi-particle peak in the spectrum, and an artificial
ferromagnetic solution is found. Reliable results can only be
expected for smaller $n$. Here, good overall agreement with QMC
data is found regarding the spectral density as well as the
static susceptibility.

Considerable improvement upon the MAA and the EHA is achieved for
the description of spontaneous ferromagnetic order on a $d=\infty$ 
fcc-type lattice. For fillings up to quarter filling and for weak 
and moderate interaction strength $U$, the IAA is able to predict 
the magnetic properties and particularly the Curie temperatures very 
reliably. Also the results for strong interactions $U/t^\ast \le 16$ 
at $n=0.5$ seem to be still plausible. 
On the other hand, the approximation breaks down for higher fillings
$n \gtrsim 0.5$. The MAA yields a clearly too high $T_C$ which 
is due to an underestimation of quasi-particle damping. The EHA 
is as reliable as the IAA at low fillings where it predicts a 
Curie temperature consistent with the QMC result. However, for 
$n \gtrsim 0.3$ the Curie temperatures are too low, 
and a paramagnetic 
ground state is found for $n > 0.51$ (at $U=4$). Thus the EHA breaks 
down at a considerably lower filling compared with the IAA.

In our opinion this is due to the effect of a higher-order
correlation function $B_{-\sigma}$ which is of special importance
for several reasons: Inclusion of $B_{-\sigma}$ is necessary to
ensure the correct high-energy behavior of the theory as well as
the correct moments of the spectral density (up to $m=3$). This
is the main condition to recover the exact strong-coupling 
results of Harris and Lange. For $U\mapsto \infty$ and in the
ferromagnetic phase, the correlation function $B_{-\sigma}$ results
in a {\em spin-dependent} shift $n_{-\sigma} B_{-\sigma}$ of the
center of gravity of the lower Hubbard band. Thus the stability
of ferromagnetic order is intimately related to the effective
band shift. It is known that ferromagnetic order cannot be found 
at all or only under extreme circumstances within the Hubbard-I or 
the alloy-analogy solution where $T_0$ appears in place of 
$B_{-\sigma}$ in the high-energy expansion coefficients of the
self-energy.

The apparent importance of the band shift for magnetic order
provokes the question whether it is possible to construct 
analytical approaches that on the one hand are exact up to the
$m=3$ moment and on the other avoid the deficiencies of the IAA
concerning its low-energy behavior. For example, up to now there
is no {\em conserving} theory to our knowledge that correctly
accounts for the $m=3$ moment.
\\

{\center \bf \noindent ACKNOWLEDGEMENT \\ \mbox{} \\} 

Support of this work by the Deutsche Forschungsgemeinschaft within 
the Sonderforschungsbereich 290 (``Metallische d\"unne Filme: 
Struktur, Magnetismus und elektronische Eigenschaften'') is
gratefully acknowledged.
\\

\small
\baselineskip3.4mm

------------------------------------------------------------------


\begin{thebibliography}{10}
\bibitem{tastdisc}
This differs from the choice of $t^\ast=t/\sqrt{2Z}$ for
the hyper-cubic lattice. For the fcc lattice $t^\ast=t/\sqrt{Z}$ 
and $T_{\langle ij \rangle}>0$ 
has been preferred here to be consistent
with the results of Refs.\ \cite{VBH+97,Ulm98}.
\end{thebibliography}
\end{document}